\documentclass[aps,prb,preprint,superscriptaddress]{revtex4-2}
\usepackage{graphicx}  
\usepackage{dcolumn}   
\usepackage{bm}        
\usepackage{amssymb}
\usepackage[english]{babel}
\usepackage{booktabs}
\usepackage{array}
\usepackage{caption}
\usepackage{subcaption}
\usepackage{hyperref}
\hypersetup{
	colorlinks=true,
	linkcolor=green,
	filecolor=magenta,      
	urlcolor=blue,
}
\begin{document}
\title{ Pressure induced emission enhancement and bandgap narrowing: experimental investigations and first principles theoretical simulations on a model halide perovskite}

\author{Debabrata Samanta}
\affiliation{National Centre for High Pressure Studies, Department of Physical Sciences, Indian Institute of Science Education and Research Kolkata, Mohanpur Campus, Mohanpur 741246, Nadia, West Bengal, India.}

\author{Sonu Pratap Chaudhary}
\affiliation{Department of Chemical Sciences, and Centre for Advanced Functional Materials, Indian Institute of Science Education and Research (IISER) Kolkata, Mohanpur-741246, India.}

\author{Bishnupada Ghosh}
\affiliation{National Centre for High Pressure Studies, Department of Physical Sciences, Indian Institute of Science Education and Research Kolkata, Mohanpur Campus, Mohanpur 741246, Nadia, West Bengal, India.}

\author{Sayan Bhattacharyya}
\affiliation{Department of Chemical Sciences, and Centre for Advanced Functional Materials, Indian Institute of Science Education and Research (IISER) Kolkata, Mohanpur-741246, India.}

\author{Gaurav Shukla}
\affiliation{Department of Earth Sciences and National Centre for High Pressure Studies, Indian Institute of Science Education and Research Kolkata, Mohanpur Campus, Mohanpur 741246, Nadia, West Bengal, India.}

\author{Goutam Dev Mukherjee}
\email [Corresponding author:]{goutamdev@iiserkol.ac.in}
\affiliation{National Centre for High Pressure Studies, Department of Physical Sciences, Indian Institute of Science Education and Research Kolkata, Mohanpur Campus, Mohanpur 741246, Nadia, West Bengal, India.}
\date{\today}

\begin{abstract} 
We report high-pressure photoluminescence, Raman scattering, and x-ray diffraction measurements on a lead-free halide perovskite $Cs_3Sb_2Br_9$. At about 3 GPa, an electronic transition manifests itself through a broad minimum in linewidth, a maximum in the intensity of $E_g$, $A_{1g}$ Raman modes, and the unusual change in the $c/a$ ratio of the trigonal lattice. The large compressibility and observed Raman anomalies indicate to a soft material with strong electron-phonon coupling. The observed below bandgap broadband emission in the photoluminescence measurement indicates the recombination of self-trapped excitons. The initial blueshift of the photoluminescence peak reinforces itself to the redshift at around 3 GPa due to the change in the electronic landscape. A first order trigonal to a monoclinic structural transition is also seen at 8 GPa. The first-principles density functional theory (DFT) calculations reveal that the electronic transition is associated with direct-to-indirect bandgap transition due to changes in the hybridization of $Sb-5s$ and $Br-4p$ orbitals near the Fermi level in the valence band. The experimentally observed Raman modes are assigned to their symmetry using the density functional perturbation theory. In addition, the DFT calculations predict a 27.5\% reduction of the bandgap in the pressure range 0-8 GPa. 
	
\end{abstract}

\maketitle

\section*{I. INTRODUCTION}
The search for energy-efficient materials has gained momentum in recent times, to use with green and renewable energy sources. In this process, the synthesis of the materials with improved efficiency in photovoltaic and optoelectronic properties has generated huge interest in the interdisciplinary field of science. Therefore, it is important to understand not only the fundamental interactions in these materials to predict new novel materials with improved properties but also their behavior under different external conditions. Pressure has been acknowledged as an important clean parameter in this direction. During this search with an emphasis on easy to synthesize and low cost, lead halide perovskites have shown  several outstanding properties such as tunable bandgap, large absorption coefficient, low exciton binding energy, high electron and hole mobility leading to good solar material, but the toxicity and poor stability have prevented them from commercial applications \cite {Zha}. In order to overcome both the toxicity and stability issues, different types of bismuth and antimony-based halide perovskites have been reported.  Amongst several candidates, low-toxic antimony and bismuth-based halide perovskites of the family  $A_3B_2X_9$ ($A$: Cs, Rb; $B$: Sb or Bi; and $X$: halogen), which are stable under atmospheric conditions have shown good promise \cite{Wu,Zhang, Liu, Zheng, Jian, Hoefler, Pradhan, Ma}. The $Cs_3Sb_2Br_9$, a member of the aforementioned family, is a superior candidate for solar cells owing to its direct bandgap, high absorption coefficient ($>10^5/cm$), high photoluminescence quantum yield (PLQY), and small effective mass ($m_e=0.34, m_h=0.42$) \cite{Zhang, Liu, Zheng}. Several studies have been published on $Cs_3Sb_2Br_9$, but mainly focussing on different types of synthesis processes \cite{Zhang, Zheng, Ma} to improve optoelectronic and photovoltaic properties. The temperature-dependent photoluminescence (PL) study has revealed exciton binding energy of 530 meV and 56.8 meV in the cases of $Cs_3Sb_2Br_9$ QD and $Cs_3Sb_2Br_9$ single crystal, respectively  \cite{Jian, Liu}, showing an important correlation between the structure and optoelectronic properties. But it is noteworthy that the wide bandgap (2.26-2.65eV) \cite{Liu, Zheng, Jian} of the $Cs_3Sb_2Br_9$ prevents its application in the optoelectronic field. It is extremely difficult to overcome the above issue in a traditional way such as chemical manipulation \cite{Akkerman, Manna} and different processing conditions \cite{Zhang, Zheng, Ma}. The indirect to direct bandgap transition, enhancement of PLQY, and increase of absorption cross-section have been observed by chemical manipulation \cite{Byang, Luo}. On the other hand, different processing conditions increase the carrier lifetime \cite {Zheng} and also improve PLQY \cite{Ma}. Therefore, as mentioned earlier, it is important to analyze and understand the fundamental interactions and the correlation between the structure and the electronic properties. To overcome the problem an alternative and efficient way is to use pressure as a parameter and study the physical properties at high densities.  Applications of hydrostatic pressure have been found to improve the optical, and the electronic properties of the halide perovskites by tuning interatomic distance without changing the material composition\cite{Wu, Zhang, Samanta, Nagaoka, Zma, Lv, Fu, Li}. Application of the pressure on halide perovskites has revealed a variety of pressure-induced phenomena such as structural phase transition \cite{Samanta, Zma, Fu}, semiconductor to metal transition \cite{Wu}, bandgap optimization \cite{Wu}, emergence of photoluminescence \cite{Samanta, Zma}, crystalline to amorphous transitions \cite{Szafranski}, etc. 
The narrowing of the bandgap under pressure is also observed in bismuth and antimony-based halide perovskites. 
The increase of hybridization of atomic orbitals driven by the reduction of  bond length and bridging bond angle  leads to narrowing of the bandgap, as seen in zero-dimensional (0D) $Cs_3Bi_2I_9$ \cite{Zhang}. Ting Geng et al. have reported a 33.7$\%$ reduction in the bandgap under high pressure in two-dimensional (2D) $Cs_3Sb_2I_9$ nanocrystals. Furthermore, first-principles calculations reveal that narrowing of bandgap is driven by the orbital interactions associated with the distortion of the $Sb–I$ octahedral network upon compression \cite{Geng}. Even after these many studies, the phase stability of bulk $Cs_3Sb_2Br_9$ under pressure, which is important during applications, remains unexplored. Therefore, we have focused on structural, optical, and electronic properties of bulk $Cs_3Sb_2Br_9$ crystal under pressure.

In the present work, we report a combination of experimental and theoretical studies involving photoluminescence, Raman scattering, powder x-ray diffraction (XRD) measurements, and first-principles calculations as a function of pressure. The $Cs_3Sb_2Br_9$ undergoes two phase transitions upon hydrostatic compression up to about 14 GPa, confirmed by Raman and XRD measurements. An enhancement in PL intensity is observed at  1.4 GPa, which is related to an increase of  population of excitons and absorption coefficient  at the excitonic energy level with an increase in pressure. In addition, the DFT calculations predict a 27.5\% reduction of the bandgap in the pressure range 0-8 GPa.

\section*{II. Experimental section}
Crystalline powder of $Cs_3Sb_2Br_9$ is prepared by acid precipitation method\cite{Samanta}. High pressure experiments are carried out using a piston-cylinder type diamond anvil cell (DAC) of culet size 300 $\mu m$. A steel gasket of thickness 290 $\mu m$ is preindented to a thickness of 55 $\mu m$ by squeezing it in between two diamonds and then a 100  $\mu m$ hole is drilled at the center of the indented portion. The top of the lower diamond together with the central hole of the gasket act as the sample chamber. The sample along with a pressure marker is loaded in the sample chamber. We use 4:1 methanol ethanol mixture as a pressure transmitting medium (PTM) to maintain hydrostatic conditions inside the sample chamber.

High-pressure Raman scattering and PL measurements are carried out using the Raman spectrometer (Monovista from SI GmBH) in the backscattering geometry. 532 nm and 488 nm laser lines are used to excite the sample for Raman and PL experiments, respectively.  An infinity-corrected 20X objective is used to focus the incident radiation beam and also to collect the scattered radiations. We have used a few grains of Ruby powder as the pressure marker and the pressure is calculated using the Ruby fluorescence technique \cite{Mao}. XRD experiments are performed at the XPRESS beamline in the Elettra synchrotron radiation source. A 20 $\mu m$ diameter monochromatic x-ray of wavelength 0.4957$\AA$ is collimated to the sample for XRD measurements. Pressure is calculated using the equation of state (EOS) of the $Ag$ pressure marker \cite{Dewaele}.
We have employed DIOPTAS software \cite{Preschera} for conversion of 2D diffraction images to intensity versus 2$\theta$ plot. All the XRD data are analyzed by CRYSFIRE \cite{Shirley}, GSAS\cite{Brian}, and VESTA \cite{Koichi} software. UV-absorption spectrum at ambient conditions is recorded using a Jasco V-670 spectrophotometer.

\section*{III. Computational Details}
First-principles density functional theory (DFT) calculations are performed using a plane-wave basis set as implemented in the QUANTUM ESPRESSO \cite{Giannozzi1, Giannozzi2} software. The core and valence electron interactions are described with the ultra-soft pseudo-potentials \cite{Vanderbilt}. We have used the ultra-soft pseudo-potentials with generalized gradient approximation(GGA) \cite{Perdew} for the calculations of pressure evolution of crystal structure, electronic structure, and phonon modes. The exchange and correlation of electrons are described using Perdew-Burke-Ernzherof (PBE) \cite{Perdew} exchange-correlation functional. A shifted $3\times3\times2$ Monkhorst-Pack \cite{Monkhorst} k-mesh is chosen for the Brillouin zone summation in the total energy calculations while a shifted, much denser  $7\times7\times6$ k-mesh is used for electronic band structure and density of state calculations. The phonon calculations are carried out using the density functional perturbation theory (DFPT) \cite{Baroni}. In phonon density of states calculations, the dynamical matrix is calculated using $2\times2\times2$ q-grid and the obtained force constants are  interpolated on a $12\times12\times12$ q-grid.  The plane-wave kinetic energy cutoff of 40 Ry and the charge density cutoff of 200 Ry are taken for aforementioned calculations. The k-point grid, kinetic energy cutoff, and charge density cutoff are chosen after a careful convergence test. Structural optimization at each pressure is performed using the variable cell-shape damp dynamic method \cite{Wentzcovitch1, Wentzcovitch2} until the magnitude of Hellman-Feynman forces on each atom are smaller than $10^{-5} Ry/a.u$. In the case of phonon calculations, the lattice parameters are fixed to their experimental values. Only the internal degrees of freedom are allowed to relax.

\section{IV. RESULTS AND DISCUSSION}
\subsection*{A. Experimental}

  The synthesized $Cs_3Sb_2Br_9$ powder is characterized by synchrotron XRD  measurements at ambient conditions.  The Rietveld refinement of the XRD pattern at ambient conditions is shown in the  FIG.1(a). The atom positions of $Cs_3Bi_2Br_9$ crystal reported by F. Lazarini \cite{Lazarini} are taken as the initial structural model for the Rietveld refinement. The Rietveld refinement of the XRD pattern under the trigonal structure with the space group $P\bar{3}m1$ results in a good fit with the calculated lattice parameters: $a$=$b$=7.8964(3)$\AA$, $c$=9.6989(8)$\AA$, which agree well with previously reported literatures \cite{Peresh, Jian}. The unit cell of $Cs_3Sb_2Br_9$ crystal consists 14 atoms and it is shown in the FIG.1(b). There is one type of $Sb$ atoms, sitting at the centers of the $SbBr_6$ octahedra. Two types of $Cs$ atoms are present, which are labeled as $Cs1$ (occupy the corners of the unit cell) and $Cs2$ (inside the unit cell). The first type of $Br$ atoms ($Br1$) occupy Wyckoff position 3e and construct bridges among neighboring octahedra. The second type of $Br$ atoms ($Br2$),  sitting at Wyckoff position 6i participate to form  $SbBr_6$ octahedra.

The PL and absorption spectra at ambient conditions are shown in FIG. 2(a).  The PL spectrum shows a broad emission centered at around 1.63 eV while an exciton absorption peak is observed at around 2.66 eV.  We observe a significant Stokes-shift of about 1.03 eV. Tauc's plot \cite{Liu} is implemented to estimate the direct optical bandgap of $Cs_3Sb_2Br_9$. The linear portion of $(\alpha h \nu)^2$  {\it vs.} $h \nu$ plot is fitted by the equation: $ (\alpha h \nu)^2 =C \times (h \nu -E_g)$;  where $\alpha$ is the absorption coefficient, $\it h$ is planck's constant, $\nu$ is the frequency, $\it C$ is a proportionality constant, and $E_g$ is the direct optical bandgap. The fit yields $E_g$=2.45 eV.  To explore the origin of PL emission, we have performed  PL measurements as a function of excitation power at ambient conditions. The PL integrated intensity follows the $I\sim L^K$ law. Where $\it L$ is the excitation power and $\it K$ is a dimensionless exponent. For exciton (free and/or bound) emission, the $\it K$ value should be in between $1<K<2$ \cite{Schmidt}. Our excitation power-dependent-PL analysis reveals a linear behavior with $\it K$=1.01 (Supplementary figure FIG. S1). This suggests the excitonic nature of the observed emission. Moreover, the free exciton manifests itself in the PL spectrum with a narrow emission. The observance of broadband emission rules out the possibility of free exciton recombination. In addition, the excitonic emission is usually observed in slightly lower energy than the energy of the bandgap. Furthermore, an unusual type of exciton appears with its broadband emission at much lower energy than the bandgap. This type of exciton is called the self-trapped exciton (STE). Again, the STE emission is a well-known characteristic of the soft halide perovskites with strong electron-phonon coupling strength \cite{Lis}. Therefore, to confirm the observed emission is an STE recombination or not, detailed pressure-dependent Raman and XRD measurements are needed.

We have recorded high-pressure PL spectra of lead-free halide perovskite $Cs_3Sb_2Br_9$ using an excitation wavelength of 488 nm. FIG. 3(a,b) display PL spectra upon compression at selected pressure points. At ambient conditions, the $Cs_3Sb_2Br_9$ shows a broad emission centered at around 1.63 eV with a long tail on the high energy side. Surprisingly, both the peak position and the intensity show a drastic change with increasing pressure. The PL profile exhibits a maximum in intensity at $\sim 1.4$ GPa. At 6.9 GPa, the PL spectrum shows a broad emission similar to the ambient PL spectrum, but the intensity is found to be lower. To investigate further, the PL spectra are fitted to the Gaussian function. The deconvolution of the PL spectrum at 1.4 GPa is shown in the Supplementary figure FIG. S2. The pressure dependence of PL peak position is represented in FIG. 3(c). The PL peak shows a blueshift up to $\sim 3$ GPa followed by a redshift. This may be an indication of a change in crystal  and/or electronic structure at $\sim 3$ GPa. As shown in FIG. 3(d), the PL integrated intensity exhibits a maximum $\sim 1.4$ GPa followed by a sudden drop  $\sim 3$ GPa.  The sudden drop of PL integrated intensity $\sim 3$ GPa may have been caused by an increase in non-radiative processes. 

Under ambient conditions, we observe six Raman modes (FIG. 4(a)(bottom panel)), which are labeled as  
 $\omega_1$(30.2 $cm^{-1}$), $\omega_2$(41.1 $cm^{-1}$), $\omega_3$(67.4 $cm^{-1}$), $\omega_4$(73.3 $cm^{-1}$), $\omega_5$(181.8 $cm^{-1}$), and $\omega_6$(210.9 $cm^{-1}$). Three high-frequency Raman modes $\omega_4$, $\omega_5$, and $\omega_6$ agree well with the previously reported Raman spectrum of $Cs_3Sb_2Br_9$ single crystal under ambient conditions \cite{Liu}. Three low-frequency modes $\omega_1$, $\omega_2$, and $\omega_3$ were not observed in the previous Raman measurements as the experimental frequency range was 50-300 $cm^{-1}$ \cite{Liu}.
We have carried out high-pressure Raman scattering measurements on $Cs_3Sb_2Br_9$ up to around 21 GPa. To extract the pressure dependence of phonon frequencies, linewidth, and intensities, Raman peaks are fitted to Lorentzian functions.  FIG. 4(a)(top panel) represents Raman spectra of $Cs_3Sb_2Br_9$ at selected pressure points. Upon increasing pressure, all the Raman modes shift towards higher frequency, and eventually, they almost disappear beyond 14 GPa. The observed blueshift of all the Raman modes is related to an increase of bond strength accompanied by the reduction in the interatomic distances. The disappearance of the Raman modes is probably an indication of disordering of the crystal beyond 14 GPa (Supplementary figure FIG. S3). It is also evident from FIG. 4(a)(top panel) that a new mode $(\omega_7)$ appears above $\sim 3$ GPa. The pressure behaviour of all the Raman mode frequencies are shown in FIG. 4(b). Two low-frequency modes ($\omega_1$, $\omega_2$) show a slope change in their linear pressure behaviour at $\sim 3$ GPa and all low-frequency modes merge with the background above 7 GPa. 
The slope of $\omega_1$ mode is 2.4 $cm^{-1}/GPa$ below 3 GPa and drops to 0.4  $cm^{-1}/GPa$ above 3 GPa. Similarly, the slope of $\omega_2$ is almost two times higher below 3 GPa than that above 3 GPa. 
In general, the frequency shift of a Raman mode is dominated by anharmonic phonon-phonon interactions. The strong anharmonic phonon-phonon interaction manifests itself through softening of Raman modes with pressure \cite{Pawbake, Rajaji}. 
The pressure variation of FWHM of two low-frequency ($\omega_2$, $\omega_3$) and high-frequency  ($\omega_5$, $\omega_6$) modes is shown in FIG. 5(a). The contribution of phonon-phonon and electron-phonon interactions decide the phonon life time, which is inversely proportional to the   FWHM of Raman mode. Therefore, a study of FWHM with pressure is an indirect way to investigate the pressure dependence of anharmonic phonon-phonon and electron-phonon coupling. 
 Ideally FWHM is expected to increase with an increase in pressure. But we have observed a broad minimum in the pressure range of about 2.5 to 5 GPa. It may be noted that we have used methanol ethanol mixture as a pressure transmitting medium, which preserves the hydrostatic conditions up to about 10 GPa. Hence, the contribution of non-hydrostatic stress to the broadening of Raman modes is completely ruled out.  
The observed unusual pressure dependence of FWHM for $\omega_2$, $\omega_3$, $\omega_5$, and  $\omega_6$ modes is attributed to strong electron-phonon coupling.   Therefore, such a broad minimum can be related to either a structural or to an electronic transition in the sample.
The variation of the integrated intensity of $\omega_2$, $\omega_3$, $\omega_5$, and  $\omega_6$ mode is represented in FIG. 5(b).  We see an increase in the integrated intensity by an order of magnitude up to $\sim 3$ GPa.  An abrupt increase of the integrated intensity up to $\sim 3$ GPa indicates an enhancement of Raman scattering cross-section. In the absence of any change in chemical composition, the enhancement in Raman intensity can be due to a change in the internal polarization of the lattice. This again can happen due to a micro structural reordering.  The Raman spectrum regains the original state upon decompression, indicating the reversibility of the disorder phase.

To explore further the observed anomalies in the high-pressure Raman analysis, we have performed high-pressure XRD measurements up to $\sim 20$ GPa. All XRD patterns are shown in the Supplementary figure FIG. S4. No changes in the XRD patterns are observed up to $\sim 8$ GPa.  Drastic changes with the appearance of several new Bragg peaks are observed  above 8 GPa, suggesting a structural transition. The XRD patterns up to about 8 GPa are analyzed by Rietveld refinements. The evolution of the lattice parameters as a function of pressure (FIG. 6(a)) shows a continuous compression up to $\sim 8$ GPa. However, the $c/a$ ratio as a function of pressure shows an interesting feature (FIG. 6(b)), which increases rapidly  up to $\sim 3$ GPa (of the order of about 2\%) and then saturates. The $c/a$ ratio in a trigonal lattice can be considered as the strain in the unit cell. The increase of the $c/$a ratio up to 3 GPa indicates to an increase in the internal strain of the lattice due to the hydrostatic compression. 
In order to understand the above effect, lattice parameters ($a$ and $c$) {\it vs.} pressure data are used to calculate linear compressibility along both $a$ and $c$-axes. The linear compressibility is defined as $\kappa_{x_i}=-\frac{1}{x_i}\frac{dx_i}{dP}$, where $i$ runs from 1 to 3 ($x_1=a, x_2=b, x_3=c$) and $\it P$ is the pressure. The evolution of both $\kappa_a$ and $\kappa_c$ is shown in FIG. 6(c). $\kappa_a$ shows a much larger change in comparison to $\kappa_c$, up to about 5 GPa indicating anisotropic compressibility. Above 5 GPa, both the axial compressibilities show similar values. It should be noted that the $Cs_3Sb_2Br_9$ is a layered material. A larger change in $\kappa_a$ indicates a larger compression along the basal plane compared to the prismatic plane and can lead to a micro-structural change in the trigonal lattice. 
It is noteworthy that a similar anisotropic compressibility has been previously observed in the 0D $Cs_3Sb_2I_9$, and $Cs_3Bi_2I_9$ crystals \cite{Wu, Zhang}.
In FIG. 6(d), we have shown the pressure dependence of volume along with the equation of state (EOS) fit using $3^{rd}$ order Birch-Murnaghan EOS \cite{Dewaele}. Very small value of bulk modulus ($B_0$ = 13.1(7) GPa) and its large pressure derivative ($B'$=6.3(5)) indicate an extremely soft lattice.  
The large compressibility can be related to weaker covalent interaction between $Cs^+$ and $SbBr_6$. The obtained bulk modulus in the trigonal phase is almost similar to that for $Cs_3Bi_2Br_9(12.3GPa)$ \cite{Samanta} but about two times smaller than that for 0D $Cs_3Bi_2I_9(23.7GPa)$ \cite{Zhang} and $Cs_3Sb_2I_9(24.7GPa)$ \cite{Wu}. The presence of the soft lattice along with differential compressibility can lead to the overlap of electronic orbitals and an electronic transition. This can be a plausible explanation  for the observation of the anomalies in the Raman mode intensity and their band width.

Indexing the XRD pattern at 9.3 GPa reveals a monoclinic structure with structural parameters: 
$a =  9.0515(7)\AA, b =  7.6480(8)\AA,  c =  10.7902(11)\AA, \beta=105.709(8)^{o}$; space group: $P21/n$. In the absence of any structural model, we have carried out the Le-Bail profile refinement of the XRD pattern at 9.3 GPa as shown in the Supplementary figure FIG. S5(b). The pressure dependence of lattice parameters (Supplementary figure FIG. S6) exhibits a nonlinear behavior. Moreover, the monoclinic structure preserves up to $\sim 13$ GPa. At 15.3 GPa, a diffuse background in the $2\theta$ range 9-12 degrees along with few broad Bragg peaks appear indicating that the sample crystallinity begins to deteriorate and it continues up to 20 GPa, the highest pressure of our experiments. The EOS fit gives $B_0=99.5(1) GPa, B^{'}=10.0(3)$ for the monoclinic phase, which is much larger than that for the trigonal phase.  The discontinuity in volume vs. pressure data $\sim 8$ GPa confirms that the observed trigonal to monoclinic structural transition is first-order in nature.

 \subsection*{B. Theory}

To understand the observed Raman anomalies and the interesting feature in the $c/a$ ratio $\sim 3$ GPa, the first-principles calculations as a function of pressure are employed for the trigonal phase. The optimized lattice parameters along with their experimental values are represented in Table I. The optimized lattice parameters agree well with our experimental lattice parameters within the typical error of DFT calculations.
The GGA overestimates the lattice parameters. However, we put more emphasis on the variation of the normalized lattice parameters ($a/a_0, c/c_0$) with pressure, where $a_0, c_0$ are the lattice parameters at ambient pressure. The calculated pressure behavior of the normalized lattice parameters is consistent with our experimental results as shown in FIG. 6(a). Our DFT calculations predict the unit cell volume to be 569.8 $\AA^3$ while the unit cell volume of 523.7 $\AA^3$ is calculated from the XRD analysis, an overestimation of about 9\%.  The evolution of the calculated $c/a$ ratio as a function of pressure is compared with the experimental data in FIG. 6(b), which is well reproduced. The normalized unit cell volume obtained from both theory and experiment follows a similar trend with increasing pressure (inset of FIG. 6(d)). The third-order Brich Murnaghan EOS is used to fit the computed volume $\it vs.$ pressure data (Supplementary figure FIG. S7). The fit estimates the bulk modulus to be 10.9 GPa (The bulk modulus of 13.1 GPa is estimated from the experiment). The GGA calculations underestimate the bulk modulus by about $16.8\%$. The discrepancy in the bulk modulus is frequently observed in DFT \cite{Brik, Khein, Lee, Dliu}. 
Bulk modulus is in close relation with unit cell volume. Therefore, the discrepancy in the bulk modulus can be correlated to error introduced in the unit cell volume.

The $Cs_3Sb_2Br_9$ crystal with space group $P\bar{3}m1$ ($D_{3d}$) contains 14 atoms per unit cell. The group theory analysis predicts 42 vibrational modes at zone center, which are represented as  $\Gamma_{vib}=A_{2g}+2A_{1u}+4A_{1g}+5E_g+7A_{2u}+9E_u$. Where $A_{1g}, E_g$ modes are Raman active and $A_{2u}, E_u$ are IR-active modes. $E_g$ and $E_u$  are double-degenerate modes. Therefore, in the Raman spectrum, we can expect nine distinct Raman active modes $(4A_{1g}+5E_g)$. (Table II)The nine Raman active modes are calculated to be: $A_{1g}$(203.5 $cm^{-1}$), $E_g$(176.2 $cm^{-1}$), $A_{1g}$(102.5 $cm^{-1}$), $E_g$(85.1 $cm^{-1}$),  $E_g$(69.3 $cm^{-1}$),  $A_{1g}$(66.5 $cm^{-1}$), $A_{1g}$(38.5 $cm^{-1}$), $E_g$(36.4 $cm^{-1}$), and $E_g$(28.3 $cm^{-1}$). The visualization of the displacements of atoms associated with each modes are shown in FIG. 7. The $A_{1g}$ modes at 203.5 $cm^{-1}$ and  102.5 $cm^{-1}$ originate due to symmetric vibration of $SbBr_6$ octahedra. The vibrations of the octahedra are accompanied by the displacement of both Sb and Br2 atoms. The displacements of the Sb atoms are along the c-axis, for both the $A_{1g}$ modes at 203.5 $cm^{-1}$ and 102.5 $cm^{-1}$. 
$E_g$ modes at 176.2 $cm^{-1}$ and 85.1 $cm^{-1}$ are assigned to asymmetric vibration of $SbBr_6$ octahedra. Both Sb and Br2 atoms take part in the asymmetric vibrations. The displacements of the Sb atoms for both the $E_g$ modes at 176.2 $cm^{-1}$ and  85.1 $cm^{-1}$ are in the $a-b$ plane . The $E_g$ mode at 69.3 $cm^{-1}$ and  $A_{1g}$ mode at 66.5 $cm^{-1}$ are associated with the symmetric and the asymmetric vibrations of $SbBr_6$ octahedra, respectively. The vibrations of the $SbBr_6$ octahedra are contributed by Br2 atoms only. Mostly, the vibrations of Cs2 atoms in the $a-b$ plane and perpendicular to the $a-b$ plane are attributed to $E_g$ mode at 38.5 $cm^{-1}$ and $A_{1g}$ mode at 36.4 $cm^{-1}$, respectively. The  $E_g$ mode  at 28.3 $cm^{-1}$ mainly originates due to the rotation of $SbBr_6$ octahedra.

Our DFT calculations predict nine distinct Raman active modes but experimentally we observe only six Raman modes at ambient conditions (Table II). Two observed high-frequency modes $\omega_5$(181.8 $cm^{-1}$), $\omega_6$(210.9 $cm^{-1}$) are assigned to calculated phonon modes $E_g$(176.2 $cm^{-1}$), $A_{1g}$(203.5 $cm^{-1}$), respectively. The calculated  $A_{1g}$ (102.5 $cm^{-1}$) mode can be related to the observed $\omega_7$ mode, which appeared beyond 3 GPa. The non-observance of $\omega_7$ mode at ambient conditions can be attributed to weak Raman scattering cross-section. The observed low frequency modes are very close to each other, therefore it is difficult to identify each mode separately. However, we make a comparison between the calculated and the observed modes in the low-frequency region. Two observed  $\omega_3$(67.4 $cm^{-1}$), $\omega_4$(73.3 $cm^{-1}$) modes can be compared to $E_g$(85.1 $cm^{-1}$),  $E_g$(69.3 $cm^{-1}$),  and $A_{1g}$(66.5 $cm^{-1}$). The experimentally observed $\omega_2$(41.1 $cm^{-1}$) mode can be assigned to  either $A_{1g}$(38.5 $cm^{-1}$) or $E_g$(36.4 $cm^{-1}$). The observed  $\omega_1$(30.2 $cm^{-1}$) mode is identified to $E_g$(28.3 $cm^{-1}$) symmetry.

The pressure evolution of Raman mode frequencies obtained from both experiments and calculations is shown in FIG. 8. The pressure variation of the calculated modes agrees well with our experimental results. The phonon density of states at selected pressure values are presented in the Supplementary figure FIG. S8. The phonon density of states spreads to a higher frequency with increasing pressure. The absence of the phonon density of states in the negative frequency indicates the stability of the structure up to $\sim 8$ GPa and no obvious anharmonic phonon-phonon interactions. The calculated results are in good agreement with the results obtained from Raman scattering measurements.

The band structure of $Cs_3Sb_2Br_9$ at 0 GPa is displayed in FIG. 9(a). It is seen that the direct bandgap occurs at $\Gamma$ point and is found to be 2 eV. The calculated bandgap value is underestimated by $18\%$ . This is a well-known drawback of standard DFT calculations with GGA \cite{Borlido}, even though it captures the correct pressure variation of the bandgap. Some sophisticated functionals, such as mBJ, HLE16, and HSE06 can predict accurate bandgap \cite{Borlido}. However, we avoided band structure calculations using these functional due to high computational costs. It should be noted that the band structure calculation using HSE06 predicts the bandgap to be 2.6 eV \cite{Liu1}. The bandgap of 1.4 eV and 2.07 eV are predicted by DFT calculation using PBEsol, and HSE06 functional with taking into account the spin-orbit coupling (SOC) effect, respectively \cite{Koliogiorgos}. The  $Cs_3Sb_2Br_9$ has a direct optical bandgap in the range 2.26-2.65 eV \cite{Liu, Zheng, Jian} and it is calculated from the absorption spectrum. The estimated direct bandgap (2.45 eV) in the present work also belongs to the same range. To shed light on orbital characteristics of the conduction and valence band, we have calculated the projected density of states (PDOS) at different pressures. As shown in FIG. 9(a), at ambient pressure the strong $p-p$ interactions throughout the conduction band are accompanied by the hybridization of $ Sb-5p$ and $Br-4p$ atomic orbitals. On the other hand, typical $s-p$ interactions from hybridization of $Br-4p$ and $Sb-5s$ atomic orbitals are present throughout the valence band. The electronic band structures and PDOS at selected pressure are shown in the Supplementary figure FIG. S9 and FIG. S10, respectively. The electronic band structure calculations as a function of pressure reveal a direct to indirect bandgap transition at 3 GPa (FIG. 9(b)). The indirect bandgap occurs at $\Gamma$ point to a low-symmetry point along  $\Gamma$-to-K direction as shown by an arrow in FIG. 9(b). At 3 GPa, as shown in FIG. 9(b), the conduction band is dominated by strong hybridizations of  $ Sb-5p$ and $Br-4p$ atomic orbitals whereas the valence band is mostly associated with the hybridization of  $Br-4p$ and $Sb-5s$ orbitals. 
 The conduction band minimum (CBM) and valence band maximum (VBM) at $\Gamma$ point are denoted by CBM and VBM1, respectively. The VBM at slightly off $\Gamma$ ($\Gamma$-to-K direction) is denoted by VBM2. The CBM shows a sharp decreasing tendency up to 3 GPa followed by a pressure invariant behavior, while VBM1 and VBM2 increase monotonically with increasing pressure (FIG. 10(a)). The CBM plays a leading role up to $\sim 3$ GPa, and beyond that pressure, VBM1 and VBM2 play the leading role in the bandgap narrowing. However, the cumulative pressure behavior of CBM, VBM1, and VBM2 results in a decrease of both direct and indirect bandgap (FIG. 10(a)). The pressure evolution of both direct and indirect bandgaps are displayed in FIG. 10(b). Both direct and indirect bandgaps are found to decrease monotonically with increasing pressure. Moreover, the DFT calculations predict a $27.5\%$ reduction of the bandgap in the pressure range 0-8 GPa.

We have partially assigned Raman modes of the $Cs_3Sb_2Br_9$ crystal based on our experimental and theoretical studies. Polarized Raman measurements on single crystals can provide accurate Raman mode assignment. Noting that low-frequency modes of $Cs_3Sb_2Br_9$ have low intensity and are very close to each other. Therefore, the assignments of these modes are extremely difficult by polarized Raman measurements at ambient conditions. In that case, the polarized Raman measurement at low temperature could provide an accurate assignment of Raman modes which is beyond the scope of our experimental study. The electron-phonon interactions and anharmonic phonon-phonon interactions contribute to the phonon linewidth. The anharmonic phonon-phonon interactions do not contain any information about the electronic band since it is independent of carrier density \cite{Saha}. The non-observance of the soft mode in both experiment and theory (Supplementary figure FIG. S8) underlines, no obvious anharmonic phonon-phonon interactions. The contribution of electron-phonon interactions to phonon linewidth comes from the scattering rate of phonons by charge carriers. This rate can be calculated from the imaginary part of the phonon self-energy. The Raman linewidth is directly related to the electronic density of state and electronic band involved in the electron-phonon matrix element \cite{Saha, Gupta}. Therefore, any changes in the electronic band structure will be reflected in the FWHM of the Raman mode. The observed broad minimum in FWHM of $\omega_2$,  $\omega_3$, $\omega_5$, and $\omega_6$ modes $\sim 3$ GPa is attributed to electron-phonon coupling associated with direct-to-indirect bandgap transition, as evidenced by our electronic band structure calculations. The CB and VB are mostly associated with strong $p-p$ and $s-p$ interactions of Sb and Br atoms, respectively. Therefore, it is correlated that the $SbBr_6$ octahedra volume contraction takes part in the narrowing of the bandgap. The pressure-dependent XRD and Raman study jointly reveal the $Cs_3Sb_2Br_9$ crystal to be a soft (low bulk modulus) material with strong electron-phonon coupling. The soft material having strong electron-phonon interactions is conducive to form the self-trapped states. Hence, the observed below bandgap broad emission in the PL measurements is attributed to the phonon-assisted recombination of self-trapped excitons \cite{Zhang, Samanta, Lis, Shi}.  The population of excitons and absorption coefficient  at the excitonic energy level increase with increasing pressure, which leads to an increase in PL intensity up to around 1.4 GPa \cite{Attilio}. At 3 GPa, the blueshift of the PL  peak reinforces itself to the redshift due to a direct-to-indirect bandgap transition. The initial blueshift can be correlated to pressure induced anisotropic compressibility below 3 GPa. At above 3 GPa,  a more compact structure with no obvious anisotropy leads to the redshift of the PL peak. This highlights how the optical response is modified by the ability to contract a crystal in a particular direction.

\section{V. Conclusions}
We have investigated an electronic transition associated with direct-to-indirect bandgap transition $\sim 3$ GPa from both experiments and first-principle density functional theory calculations. High-pressure Raman measurements show a broad minimum in linewidth of $A_{1g}$ and $E_g$ modes at $\sim 3$ GPa due to unusual electron-phonon coupling. A larger compression along the basal plane compared to the prismatic plane leads to a micro-structural change in the trigonal lattice, as evidenced from XRD analysis. Self-trapped exciton manifests itself in the photoluminescence spectra through below bandgap broadband emission and the emission gets enhanced under pressure. The DFT calculation reveals that the electronic transition is associated with direct-to-indirect bandgap transition. Intriguingly, a  $27.5\%$ narrowing of the bandgap is estimated by the first-principles calculations within the GGA to an electronic exchange and correlation functional.  Furthermore, our work not only provides a relationship between optical properties and structural evolution of the $Cs_3Sb_2Br_9$ under pressure but also shows an effective way for achieving high optical response by tuning interatomic distances without changing the material composition. We also believe that this work will influence further research to understand the peculiarity of property-structure relation of the lead-free halide perovskites.

\section{Acknowledgments}
We acknowledge the financial support from the Ministry of Earth Sciences, Government of India, grant number MoES/16/25/10-RDEAS. The financial support from Department of Science and Technology, Government of India under Indo-Italy Executive Programme of Scientific and Technological Cooperation is gratefully acknowledged. The XRD measurements are carried out in the spare time of the proposal ID: 20190546. DS also acknowledges the fellowship grant supported by the INSPIRE program, Department of Science and Technology, Government of India.

\begin{table}[h]
	\caption*{Table I}{A list of optimized lattice parameters and their experimental values at ambient conditions.} 
	
	\begin{tabular}{c@{\hskip 0.5in} c@{\hskip 0.5in} c@{\hskip 0.5in} c@{\hskip 0.5in} c@{\hskip 0.5in}}
		\hline\hline 
		Approximation&Lattice parameters&Experiment& Theory& Error  \\ 
		& &($\AA$)& ($\AA$)& in $\%$  \\ 
		
		\hline 
		
		GGA &$ a$ &7.8964& 8.1334& 3 \\
		& $c $&9.6989 &9.9460 & 2.5\\
		
		\hline\hline
	\end{tabular}
\end{table}

\begin{table}[h]
	\caption*{Table II}{A list of experimental and calculated Raman mode frequencies at ambient conditions.} 
	
	\begin{tabular}{ c@{\hskip 0.7in} c@{\hskip 0.7in} c@{\hskip 0.7in} c@{\hskip 0.1in} }
		\hline\hline

		\multicolumn{2}{c}{Experimental frequencies}&
		\multicolumn{2}{c}{Calculated frequencies}\\
		\hline
		Mode&Mode frequency&Mode symmetry& Mode frequency \\ 
		& $cm^{-1}$ & & $cm^{-1}$  \\ 
		
		\hline 
		
		$\omega_1$&$ 30.2$ &$E_{g}$& 28.3  \\
		$\omega_2$&$ 41.1$ &$E_{g}$& 36.4 \\
		$\omega_3$&$ 67.4$ & $A_{1g}$& 38.5\\
		$\omega_4$&$ 73.3$ & $A_{1g}$& 66.5 \\
		$\omega_5$&$ 181.8$ &$E_{g}$& 69.3 \\
		$\omega_6$&$ 210.9$ &$E_{g}$& 85.1\\
		& &$A_{1g}$& 102.5 \\
		& &$E_{g}$& 176.2 \\
		& &$A_{1g}$& 203.5\\
		\hline\hline
	\end{tabular}
\end{table}

\begin{figure}[h]
	\centering
	\includegraphics[scale = .7]{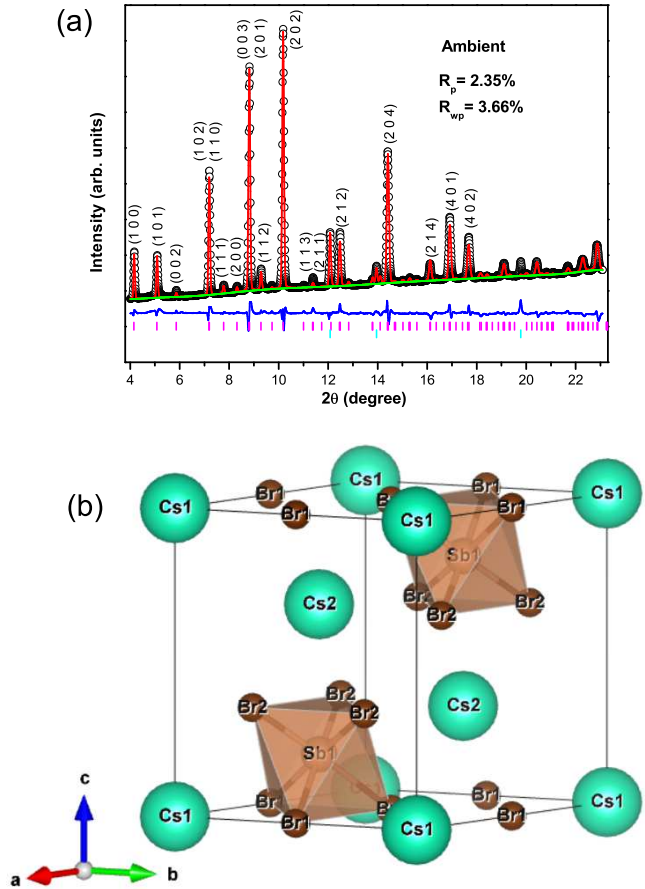}
	\vspace*{-50mm}
	\caption*{FIG. 1. (a) The Rietveld refinement of the XRD pattern at ambient conditions. The Rietveld refinement shows a good fit with lattice parameters: $a$=$b$=7.8964(3)$\AA$, $c$=9.6989(8)$\AA$; space group $P\bar{3}m1$. The black circles are experimental data. The red line shows the Rietveld fit to experimental data. The background is represented by the green line. The blue line shows the difference between experimental and calculated data. The small vertical lines show the Bragg peak of the sample (magenta) and pressure marker(cyan).(b) The unit cell of $Cs_3Sb_2Br_9$  at ambient conditions. }
\end{figure}

\begin{figure}[h]
	\centering
	\includegraphics[scale = .7]{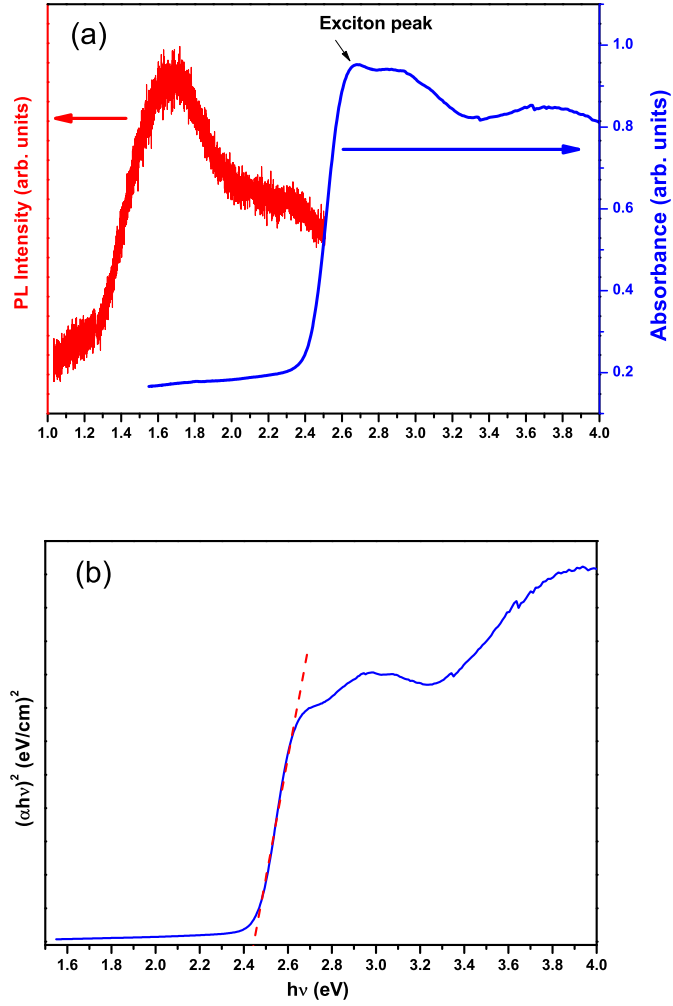}
	\vspace*{-50mm}
	\caption*{FIG. 2. (a) The PL and absorption spectra of $Cs_3Sb_2Br_9$ at ambient conditions. (b) Tauc's plot of direct bandgap semiconductor $Cs_3Sb_2Br_9$ using UV-visible spectrum at ambient conditions. Tauc's plot method estimates a direct bandgap of 2.45 eV.}
\end{figure}

\begin{figure}
	\centering
	\includegraphics[scale = .8]{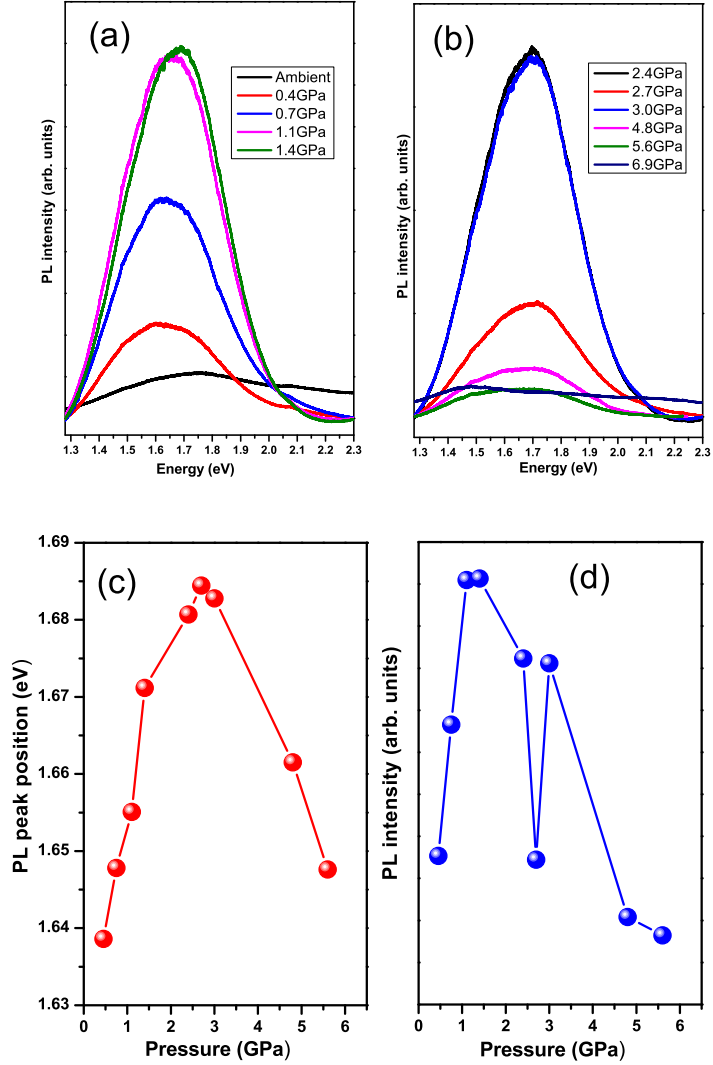}
	\vspace*{-60mm}
	\caption*{FIG. 3. (a,b) The pressure evolution of PL spectra of $Cs_3Sb_2Br_9$. The sample is excited with a laser of wavelength 488 nm. (c) PL peak position (d) integrated PL intensity as a function of pressure.}
\end{figure}

\begin{figure}
	\centering
	\includegraphics[scale = 0.7]{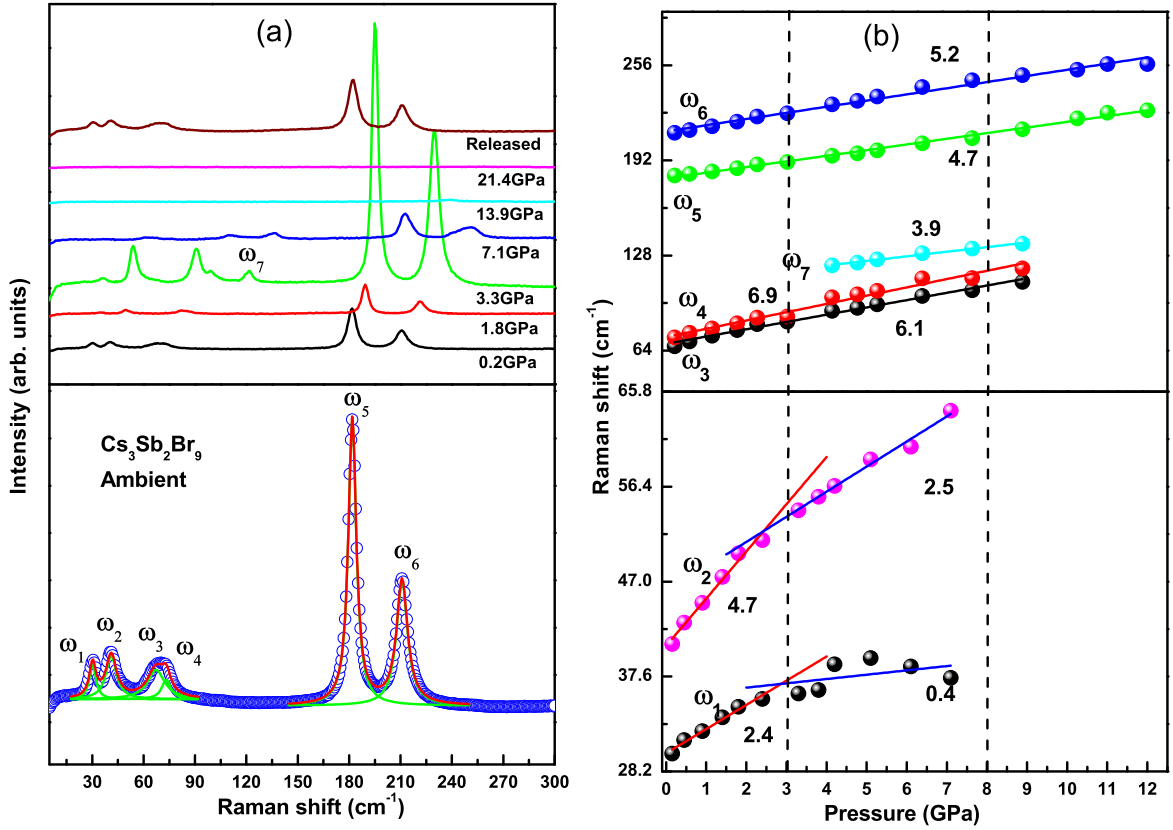}
	\vspace*{-30mm}
    \caption*{FIG. 4. (a) The deconvolution of the Raman spectrum of $Cs_3Sb_2Br_9$ at ambient conditions. 532 nm laser is used to record Raman spectra under pressure. Blue circles are experimental data. The green lines are the Lorentzian fit to experimental data. The sum of the fit is shown by red lines (bottom panel). Raman spectra of $Cs_3Sb_2Br_9$ at selected pressures (top panel). (b) The pressure dependence of Raman shift of $Cs_3Sb_2Br_9$ crystal. Solid lines represent the linear fit and the slope (in $cm^{-1}/GPa$) of the linear fit is shown alongside the lines. }
\end{figure}

\begin{figure}
	\centering
	\includegraphics[scale = 0.7]{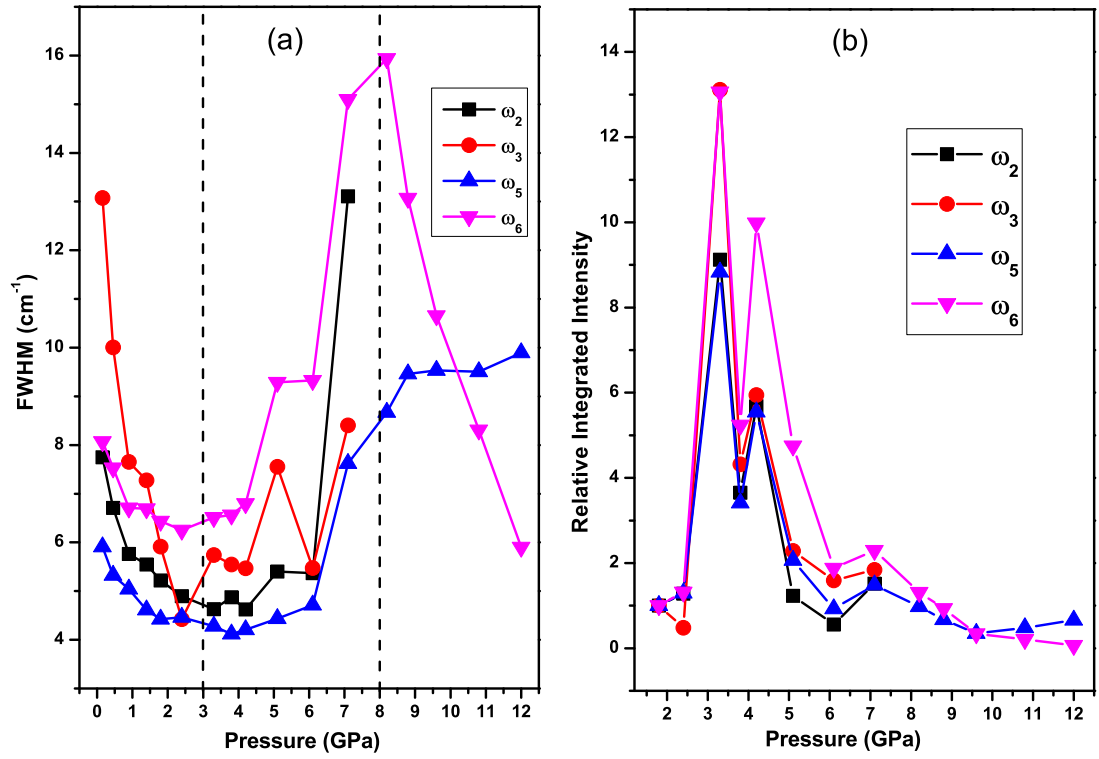}
	\vspace*{-40mm}
	\caption*{FIG. 5. The variations of (a) FWHM and (b) relative integrated intensity of $\omega_2$, $\omega_3$, $\omega_5$, and $\omega_6$ modes with pressure.}
\end{figure}

\begin{figure}
	\centering
	\includegraphics[scale = 0.7]{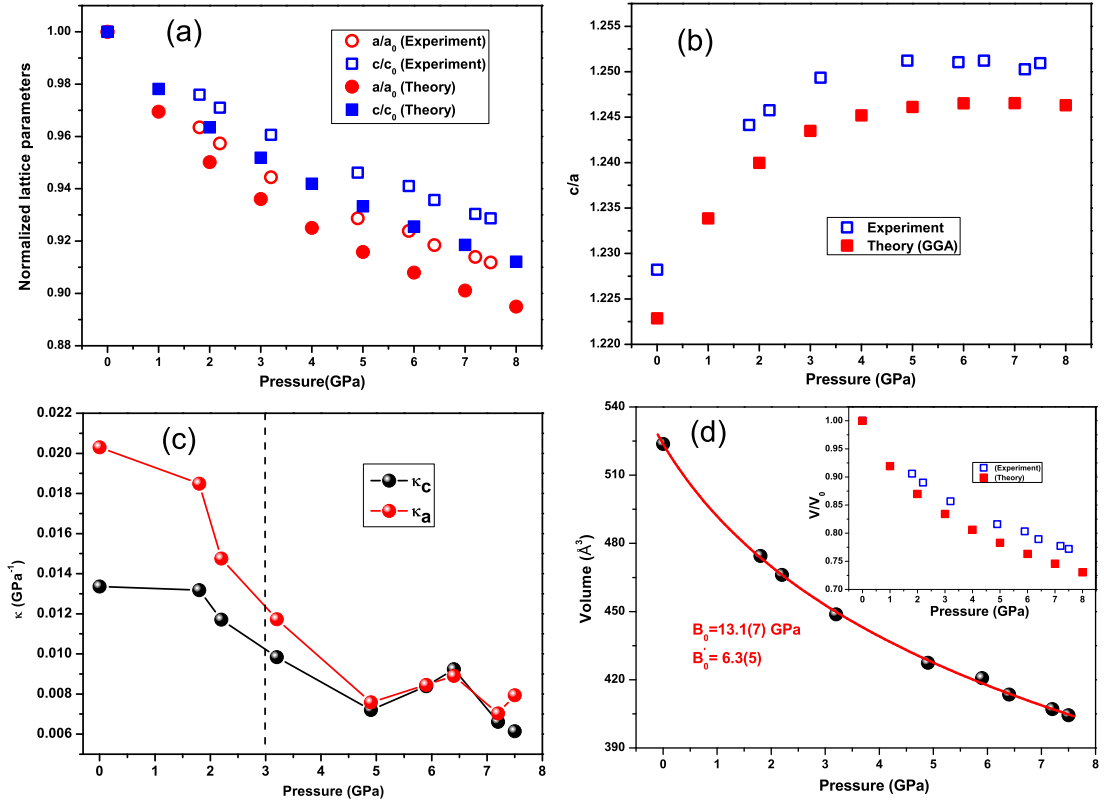}
	\vspace*{-35mm}
	\caption*{FIG. 6. The figure shows pressure behavior of structural parameters obtained from both theory and experiment. The open circles and the squares are experimental data.  The calculated data are shown by the solid circles and the squares.  (a) The pressure dependence of the normalized lattice parameters. (b) The change in the $c/a$ ratio as a function of pressure. (c) The variation of liner compressibilities as a function of pressure. (d) The evolution of the cell volume with pressure. The third-order Birch Murnaghan EOS is used to fit the experimental data. The solid line shows the fit. The pressure evolution of the $V/V_0$ is shown in the inset.}
\end{figure}

\begin{figure}
	\centering
	\includegraphics[scale = 1.5]{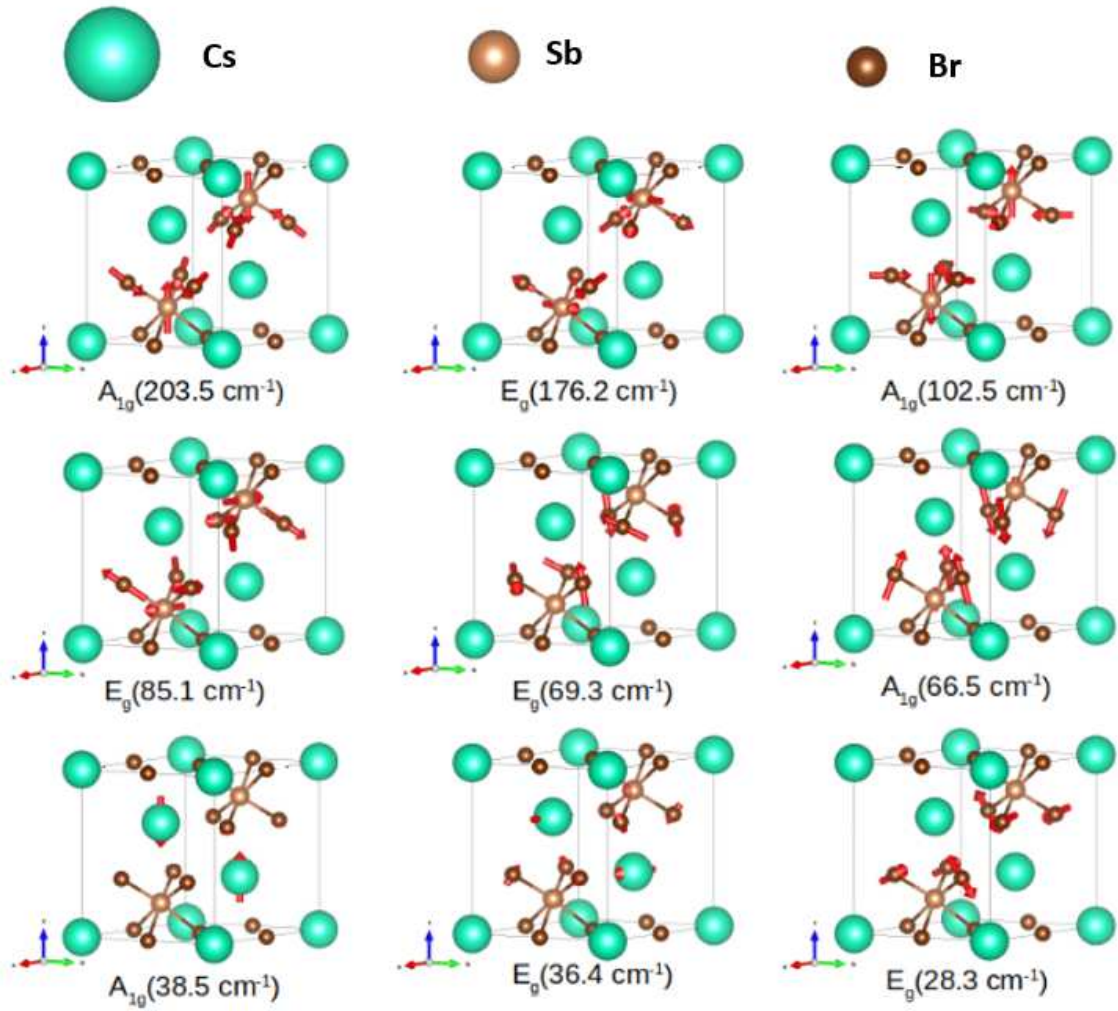}
	\vspace{13mm}
	\caption*{FIG. 7. (a) Visualization of the atomic displacements of the selected Raman modes obtained from DFPT calculations.}
\end{figure}

\begin{figure}
	\centering
	\includegraphics[scale = 0.7]{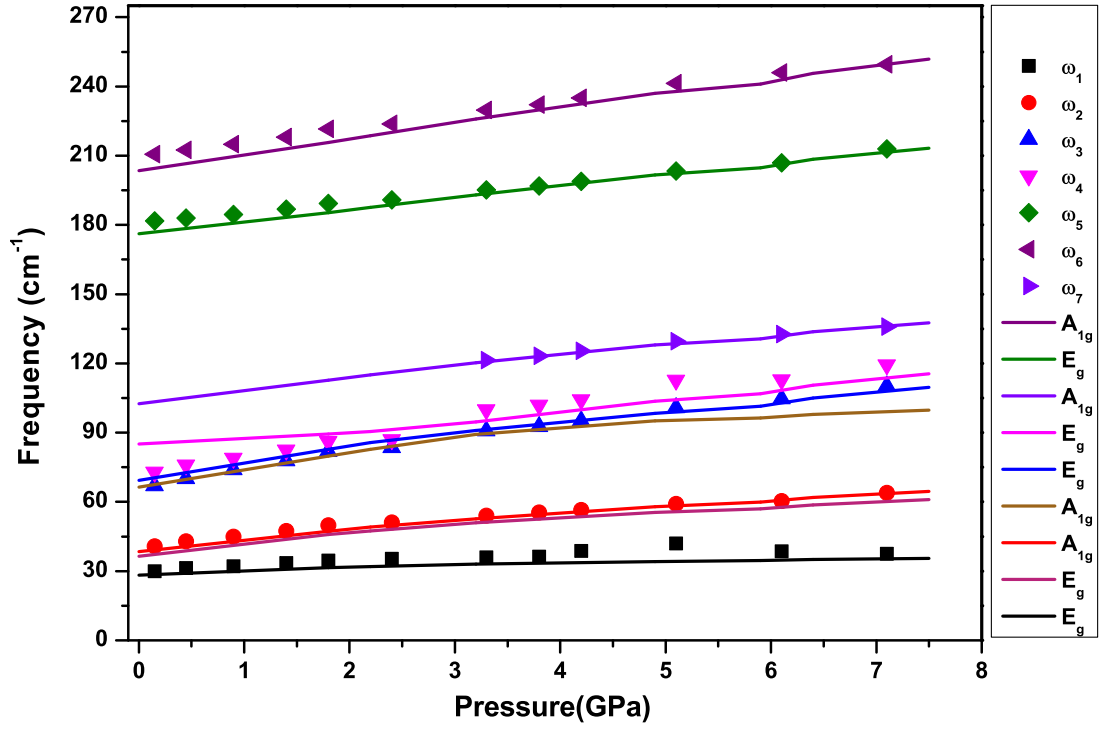}
	\vspace*{-20mm}
	\caption*{FIG. 8. (a) The pressure variation of Raman mode frequencies of $Cs_3Sb_2Br_9$ from DFT calculations and experiment. The points are experimental data. The calculated data are shown by solid lines. }
\end{figure}

\begin{figure}
	\centering
	\includegraphics[scale = 0.8]{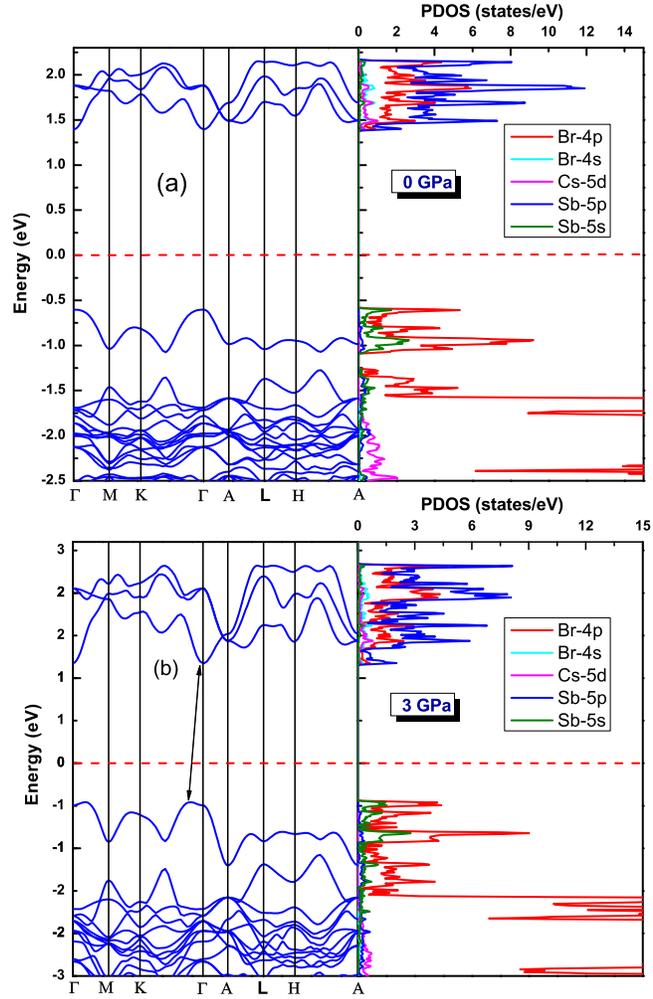}
	\vspace*{-30mm}
	\caption*{FIG. 9. (a) The Electronic band structure and projected electronic density of states at 0 GPa. (b) The Electronic band structures and projected electronic density of states at 3 GPa. The arrow in the band structure is showing the direct to indirect bandgap transition. }
\end{figure}

\begin{figure}
	\centering
	\includegraphics[scale = 0.7]{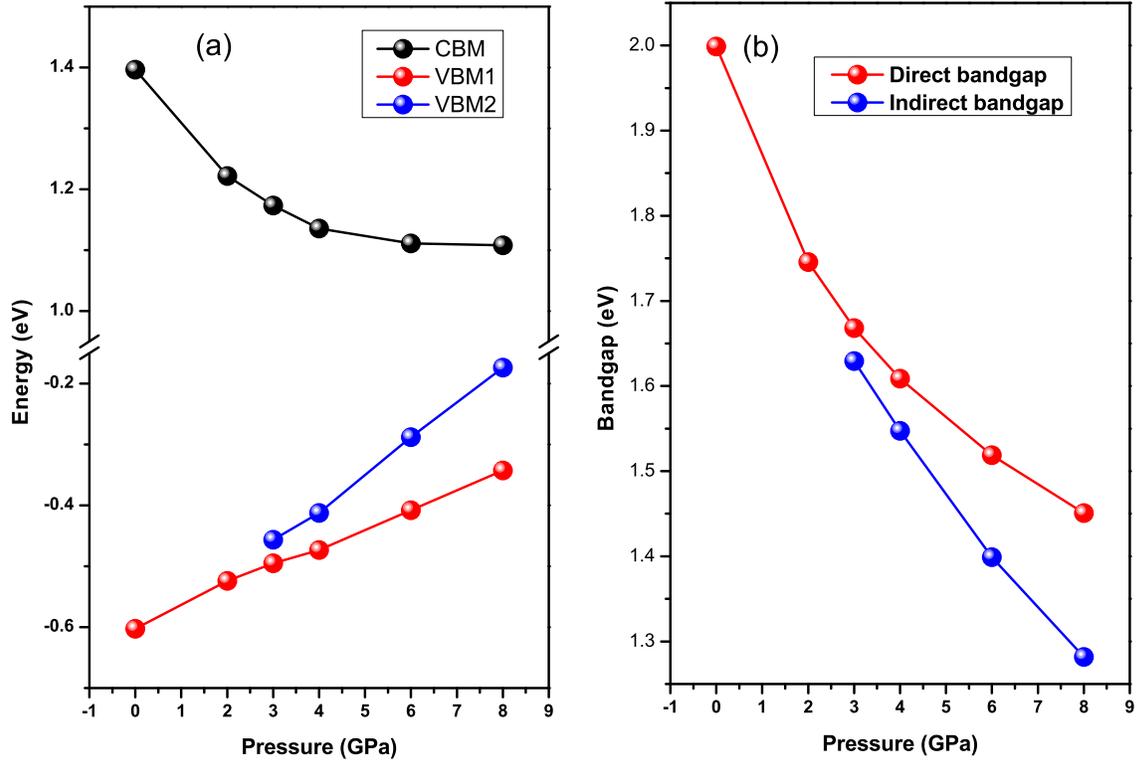}
	\vspace*{-35mm}
	\caption*{FIG. 10. (a) The pressure evolution of the VBM and CBM. (b) The evolution of both the direct and indirect bandgaps as a function of pressure.}
\end{figure}

\end{document}


\title{Supplementary Information on Pressure induced emission enhancement and bandgap narrowing: experimental investigations and first principles theoretical simulations on a model halide perovskite}

\author{Debabrata Samanta}
\affiliation{National Centre for High Pressure Studies, Department of Physical Sciences, Indian Institute of Science Education and Research Kolkata, Mohanpur Campus, Mohanpur 741246, Nadia, West Bengal, India.}

\author{Sonu Pratap Chaudhary}
\affiliation{Department of Chemical Sciences, and Centre for Advanced Functional Materials, Indian Institute of Science Education and Research (IISER) Kolkata, Mohanpur-741246, India.}

\author{Bishnupada Ghosh}
\affiliation{National Centre for High Pressure Studies, Department of Physical Sciences, Indian Institute of Science Education and Research Kolkata, Mohanpur Campus, Mohanpur 741246, Nadia, West Bengal, India.}

\author{Sayan Bhattacharyya}
\affiliation{Department of Chemical Sciences, and Centre for Advanced Functional Materials, Indian Institute of Science Education and Research (IISER) Kolkata, Mohanpur-741246, India.}

\author{Gaurav Shukla}
\affiliation{Department of Earth Sciences and National Centre for High Pressure Studies, Indian Institute of Science Education and Research Kolkata, Mohanpur Campus, Mohanpur 741246, Nadia, West Bengal, India.}

\author{Goutam Dev Mukherjee}
\email [Corresponding author:]{goutamdev@iiserkol.ac.in}
\affiliation{National Centre for High Pressure Studies, Department of Physical Sciences, Indian Institute of Science Education and Research Kolkata, Mohanpur Campus, Mohanpur 741246, Nadia, West Bengal, India.}
\date{\today}
\maketitle

\noindent Additional figures in support of the analyses carried out are presented here.

\newpage

\begin{figure}
	\centering
	
	\includegraphics[scale = 0.73]{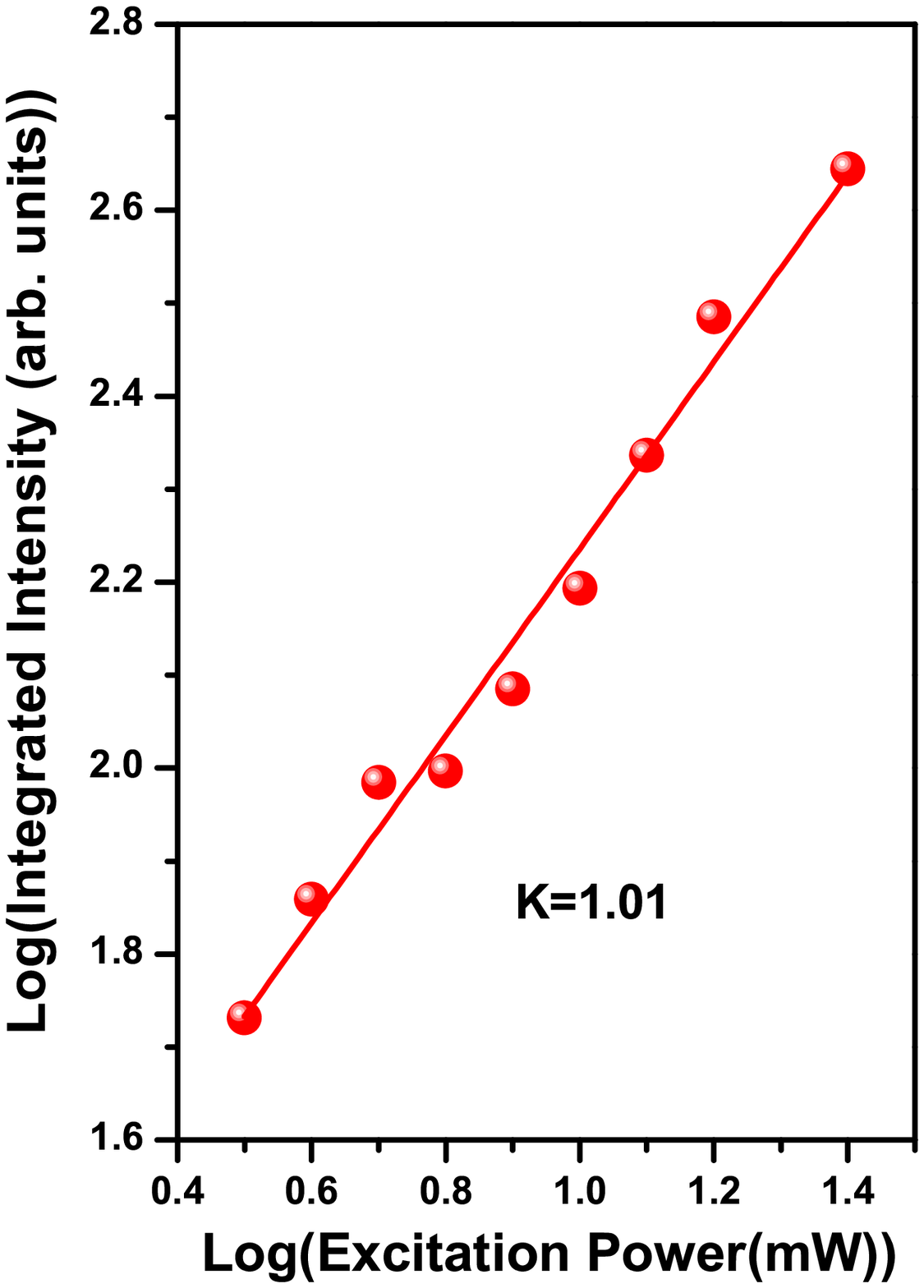}
	\vspace*{-40mm}
	\caption*{FIG. S1.  The integrated intensity (I) follows a $I \sim L^K$ law. Where K is a dimensionless constant and L is the excitation power. The integrated intensity varies superlinearly with K of the value 1.01. }
\end{figure}

\begin{figure}
	\centering
	\includegraphics[scale = 0.73]{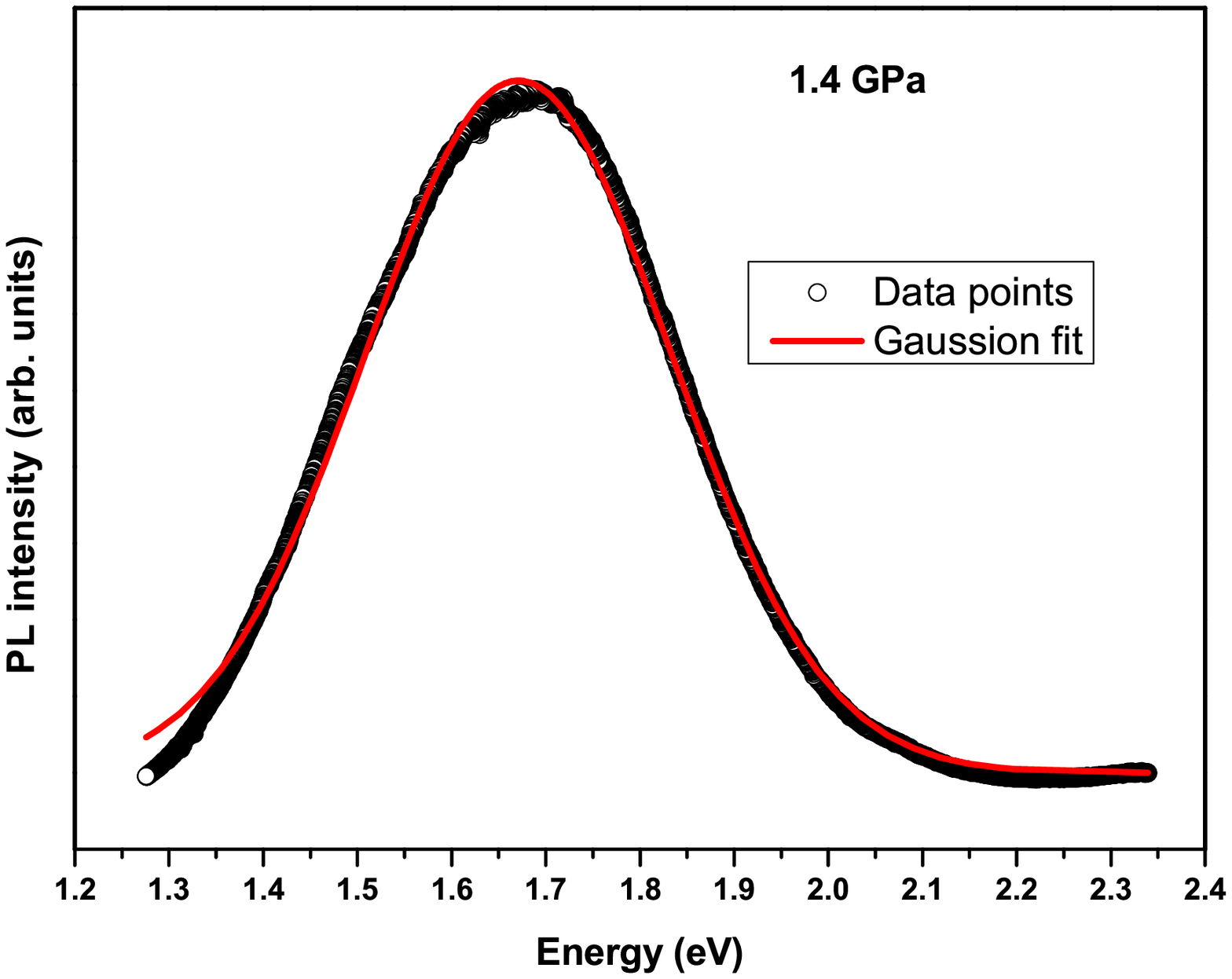}
	\vspace*{-40mm}
	\caption*{FIG. S2. The PL spectrum of $Cs_3Sb_2Br_9$ at 1.4 GPa. The dark circles are experimental data points. The red line shows the Gaussion fit to experimental data. }
\end{figure}

\begin{figure}
	\centering
	\includegraphics[scale = 0.73]{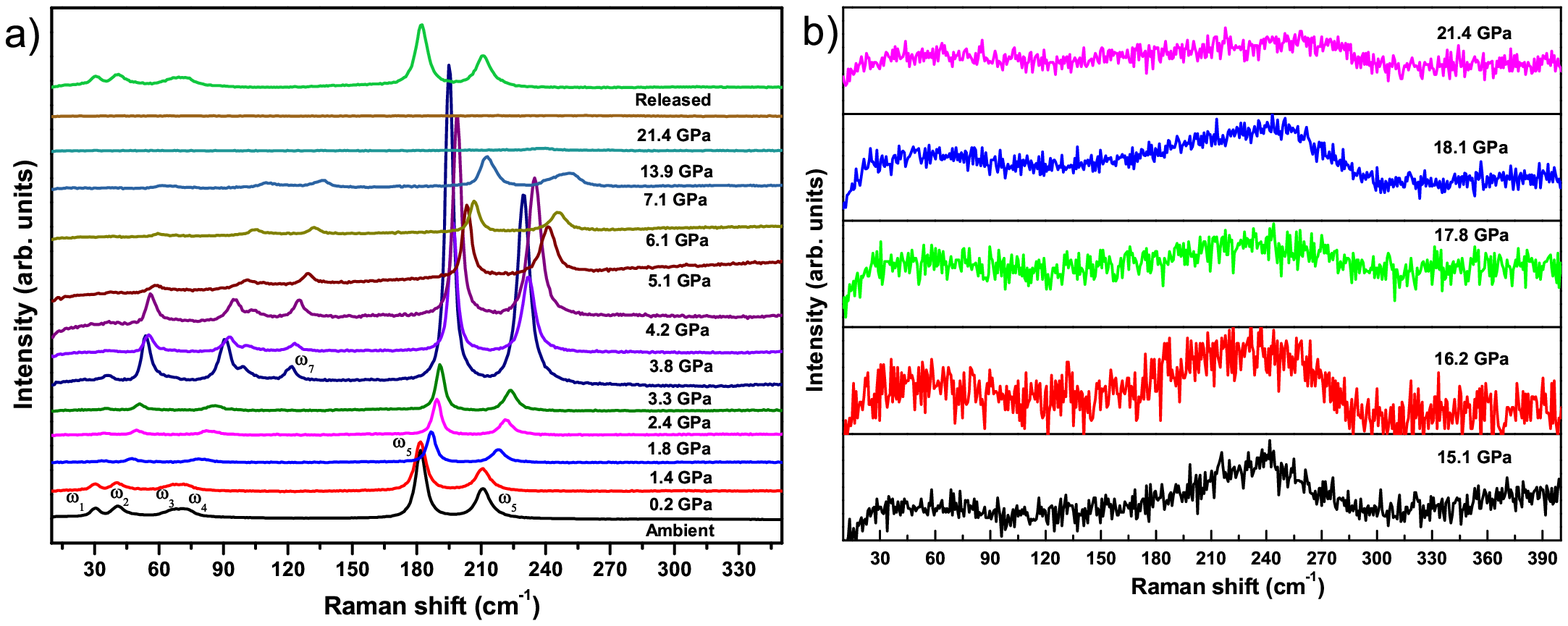}
	\vspace*{-40mm}
	\caption*{FIG. S3. Raman spectra of $Cs_3Sb_2Br_9$ at selected pressures. Raman spectra beyond 14 GPa, suggest disorder nature of the crystal.}
\end{figure}

\begin{figure}
	\begin{subfigure}[b]{0.8\textwidth}
		\centering
		\includegraphics[scale = 0.65]{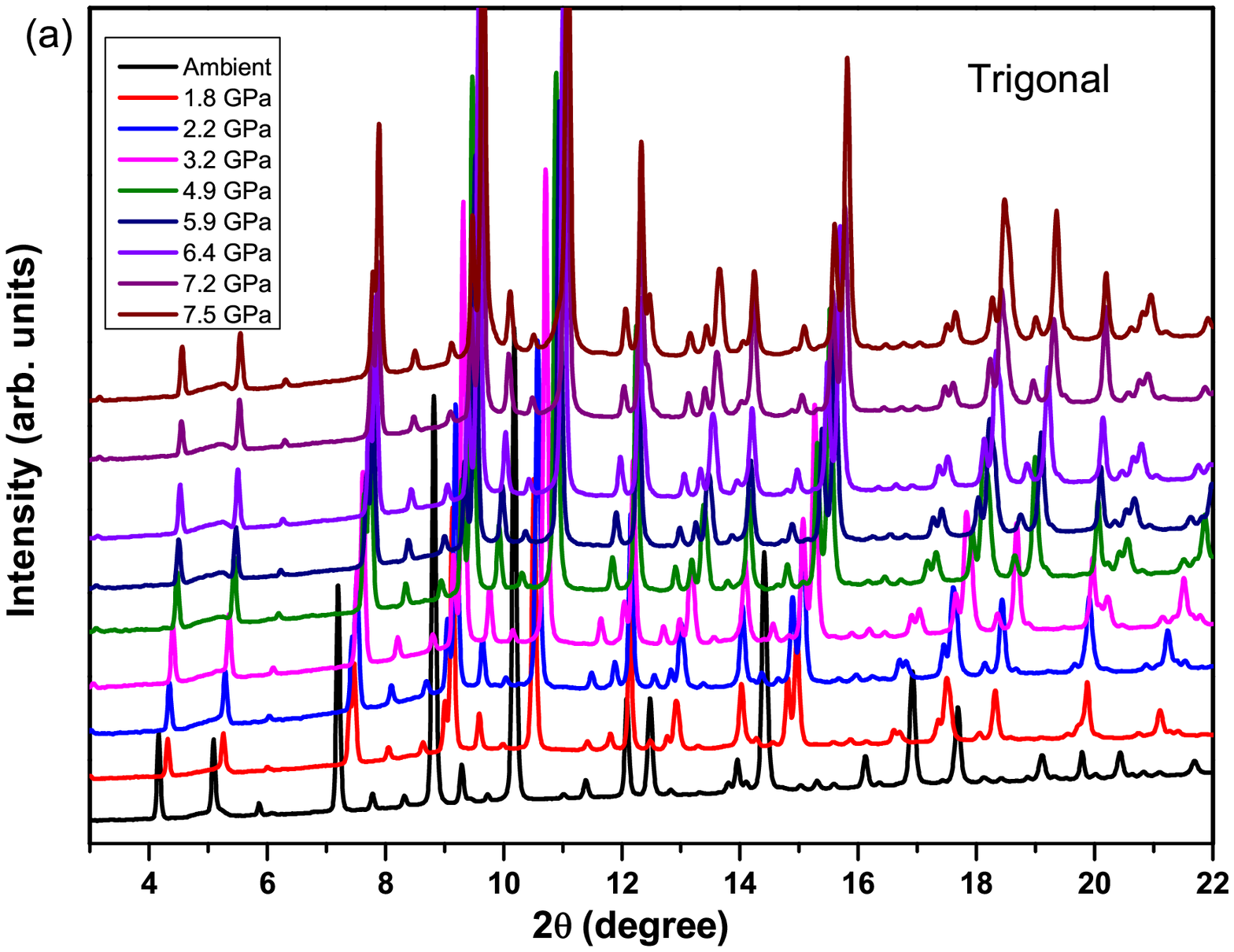}
		\vspace*{-35mm}	
	\end{subfigure}	
	\begin{subfigure}[b]{0.8\textwidth}
		\centering
		\includegraphics[scale = 0.64]{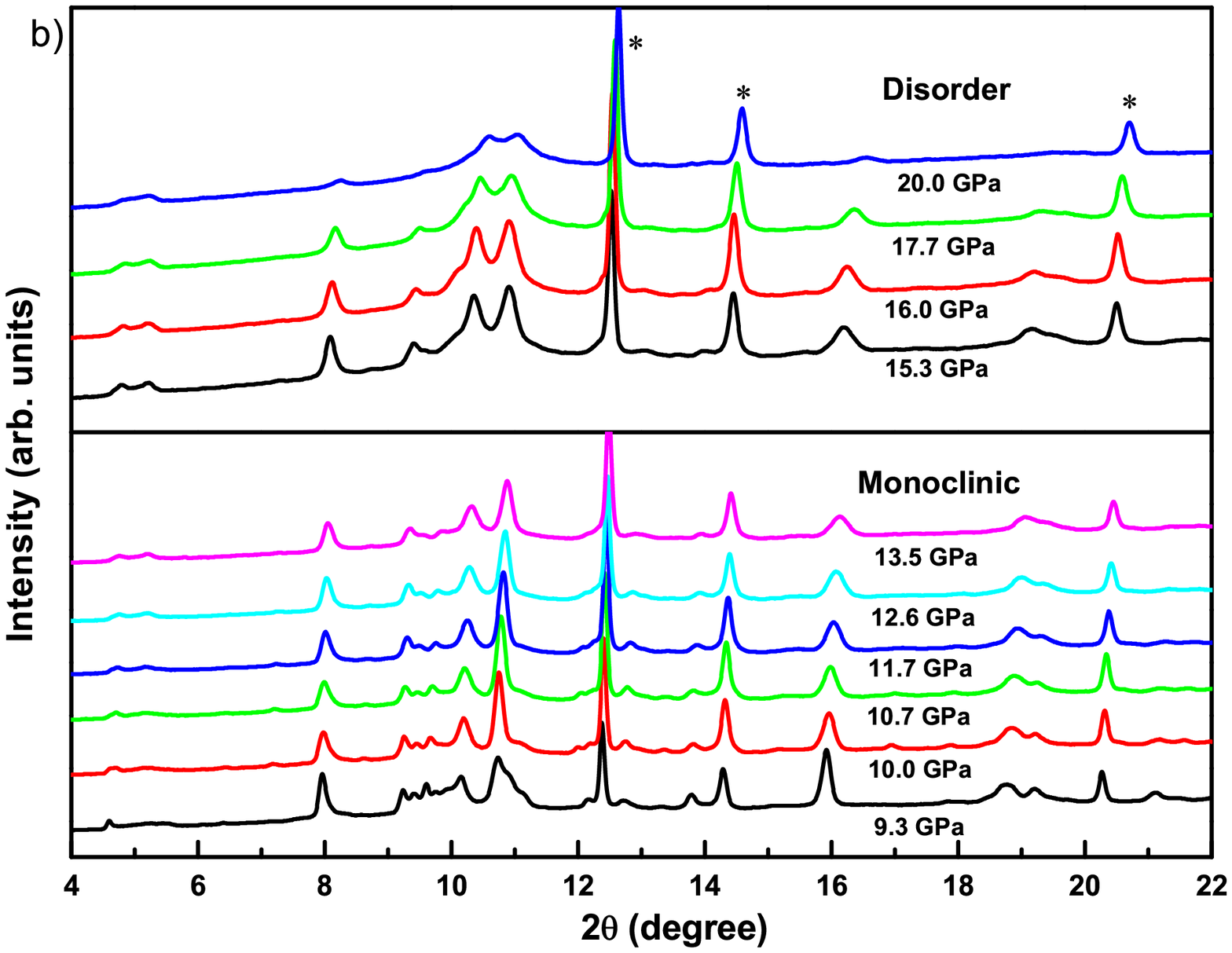}
	\end{subfigure}
	\vspace*{-25mm}
	\caption*{FIG. S4. XRD patterns at elevated pressure in the (a) trigonal phase, (b) monoclinic, and disorder phase. Asterisks (\text{*}) represent Bragg peaks of pressure marker(Ag).}
\end{figure}

\begin{figure}
	\begin{subfigure}[b]{0.8\textwidth}
		\centering
		\includegraphics[scale = 0.5]{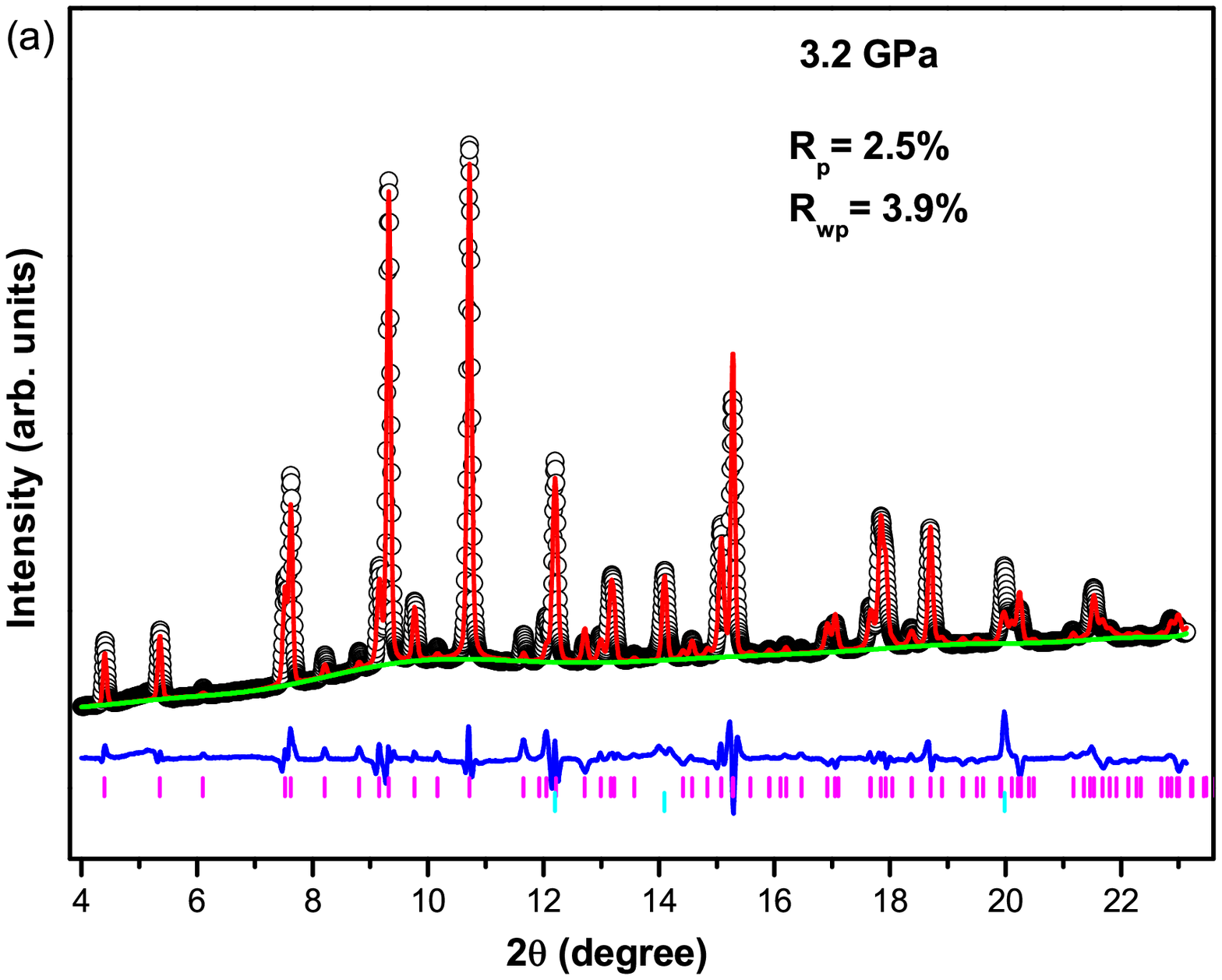}
		\vspace*{-40mm}
	\end{subfigure}	
	
	\begin{subfigure}[b]{0.8\textwidth}
		\centering
		\includegraphics[scale = 0.5]{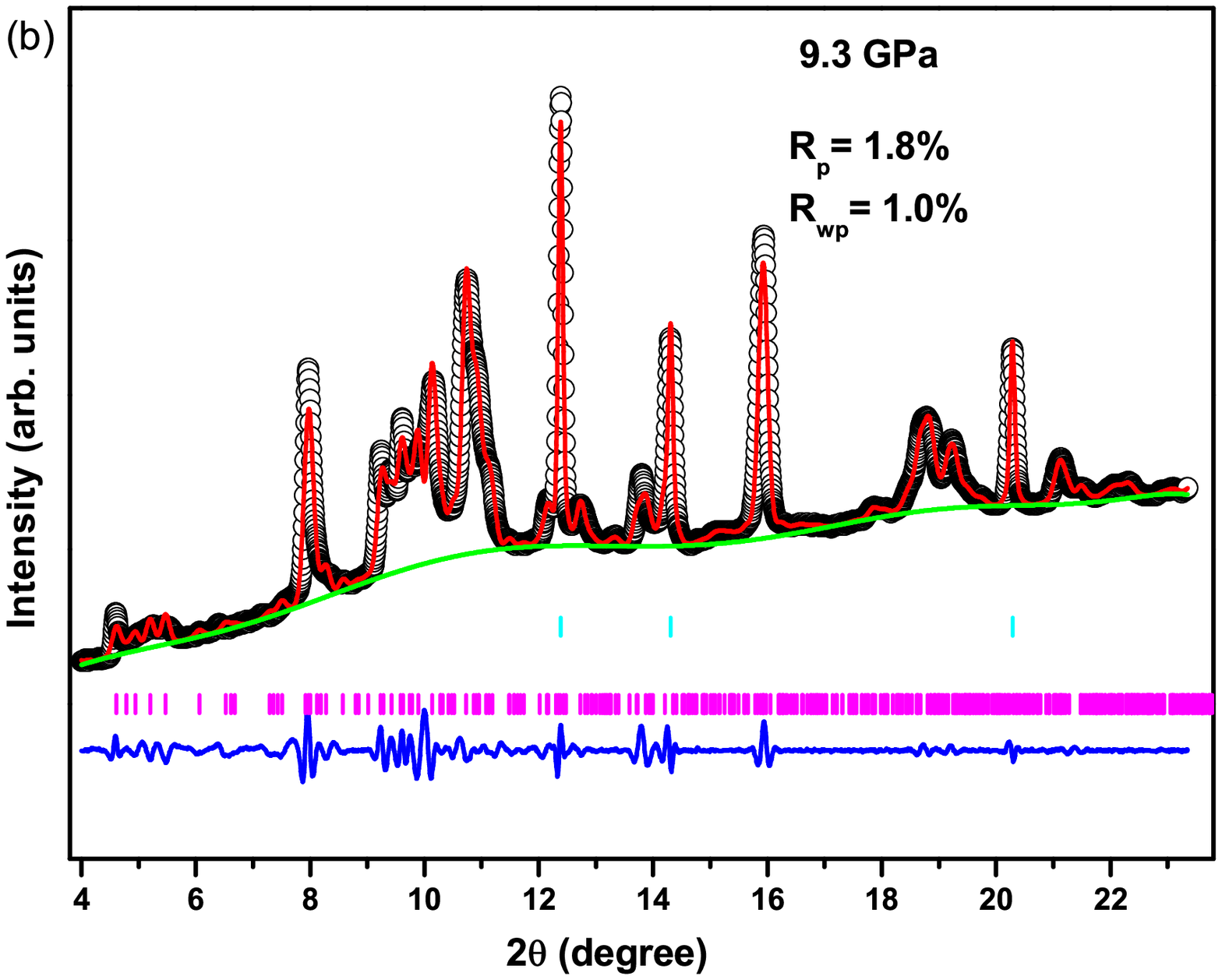}
	\end{subfigure}
	\vspace*{-25mm}
	\caption*{FIG. S5. (a)The Rietveld refinement XRD pattern at 3.2 GPa.(b) Le-Bail fit of the XRD pattern at 9.3 GPa. The black circles are experimental data. The red line shows the Rietveld fit to experimental data. The background is represented by the green line. The blue line shows the difference between experimental and calculated data. The small vertical lines show the Bragg peak of the sample (magenta) and pressure marker(cyan). }
\end{figure}

\begin{figure}
	\centering
	\includegraphics[scale = 0.73]{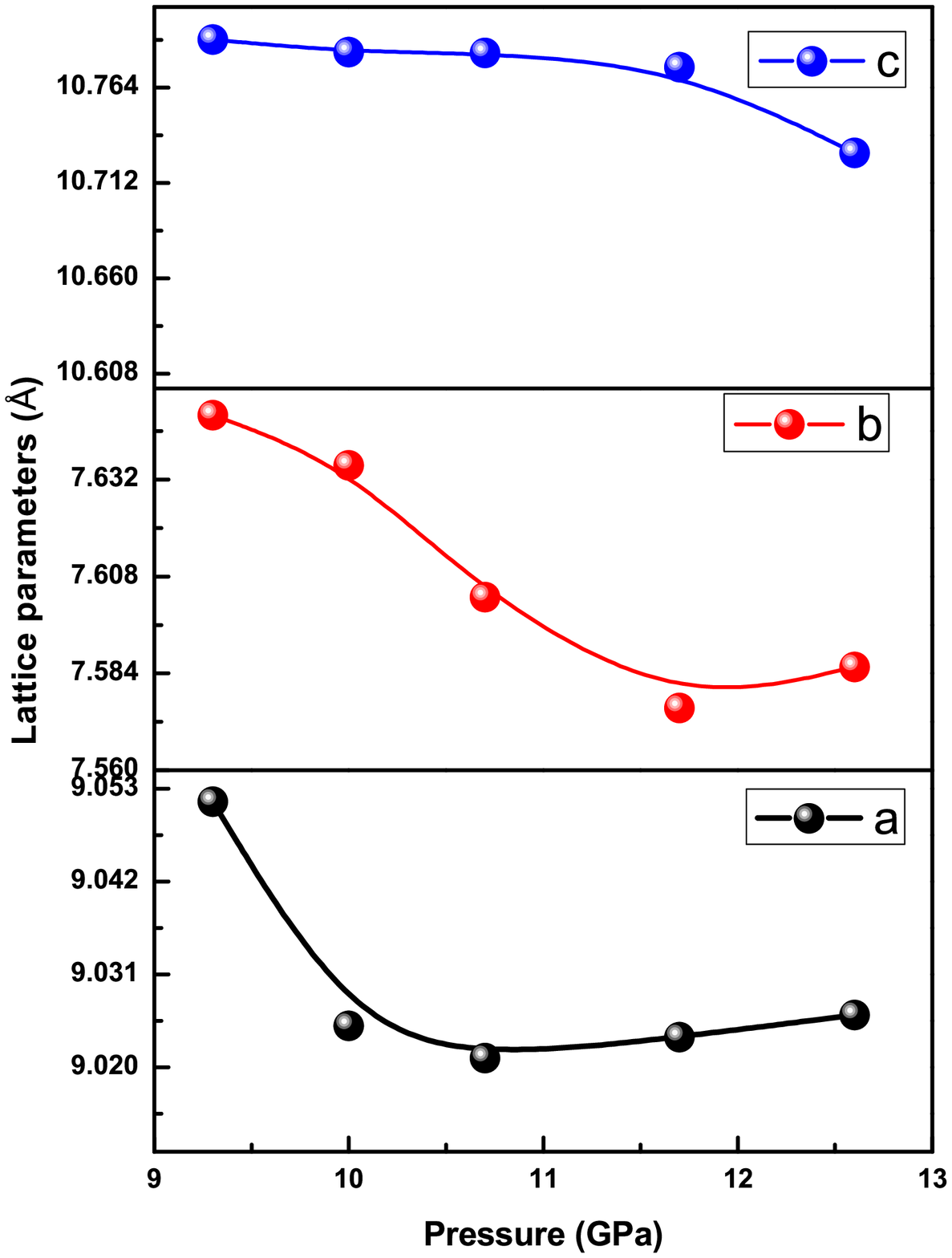}
	\vspace*{-50mm}
	\caption*{FIG. S6. The pressure evolution of the lattice parameters in the monoclinic phase.}
\end{figure}

\begin{figure}
	\centering
	\includegraphics[scale = 0.73]{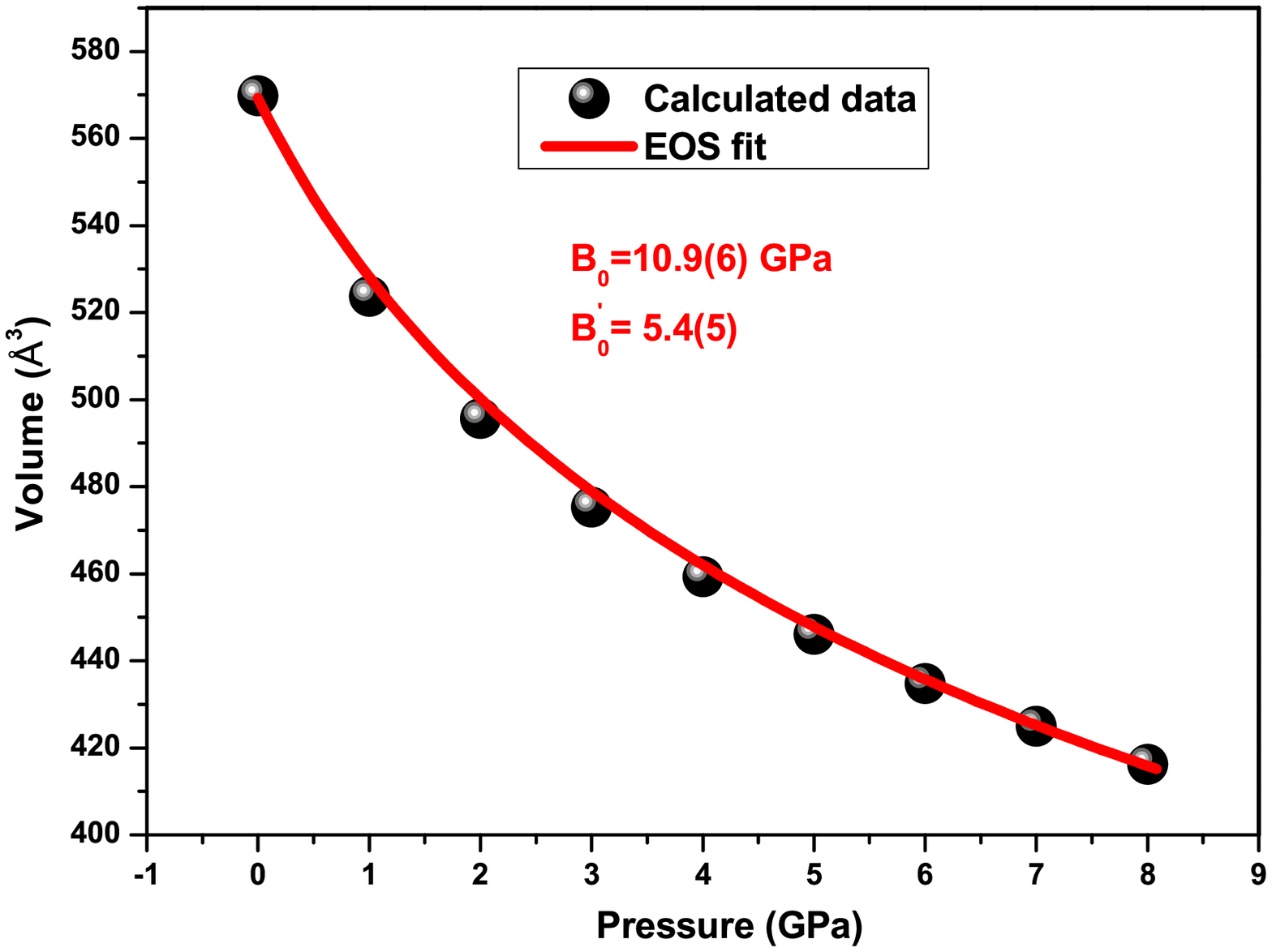}
	\vspace*{-50mm}
	\caption*{FIG. S7. The plot of computed volume vs. pressure  data. The solid spheres are calculated data. The solid line shows the third-order Birch-Murnaghan EOS fit to data. }
\end{figure}

\begin{figure}
	\centering
	\includegraphics[scale = 0.73]{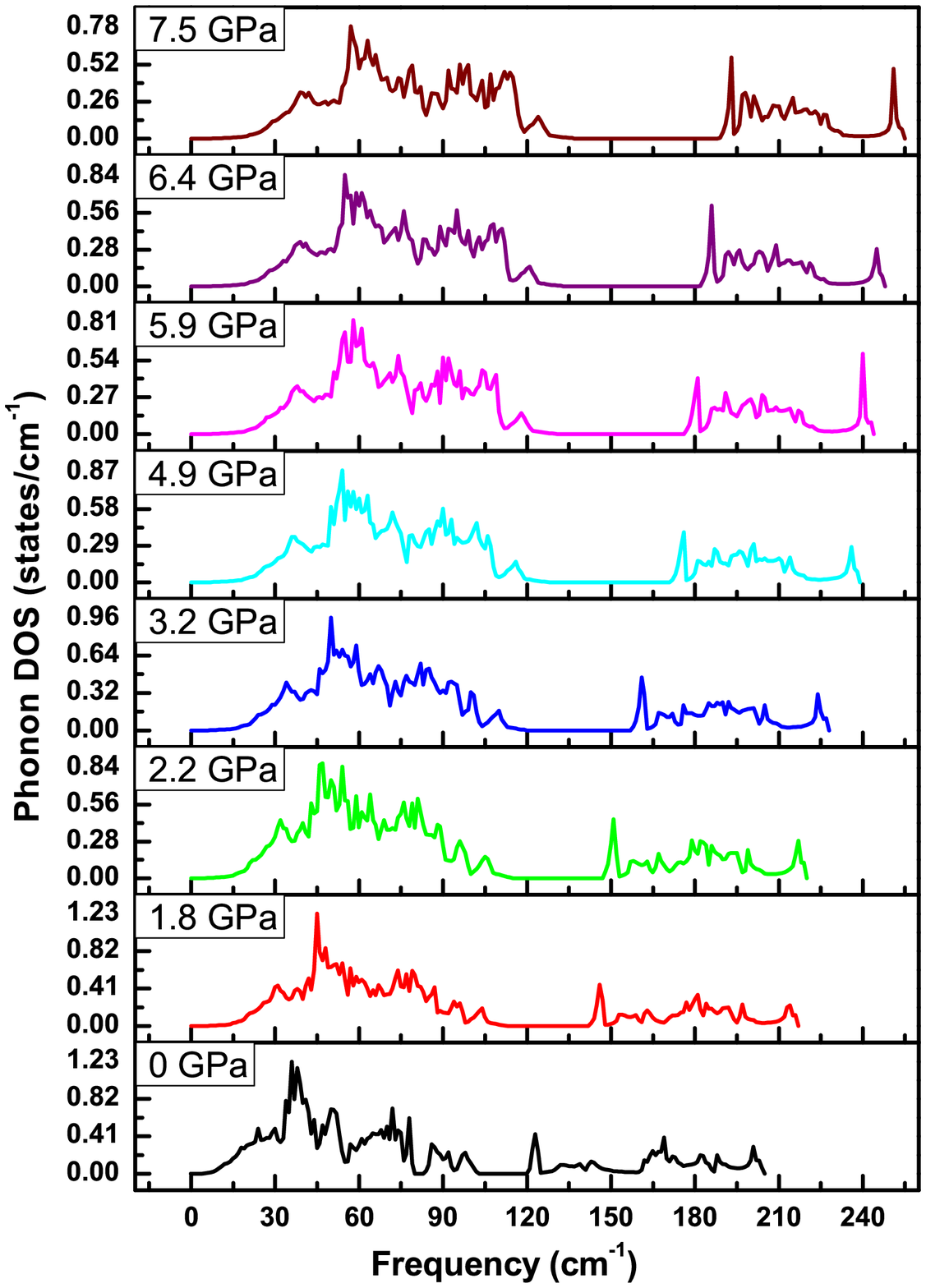}
	\vspace*{-25mm}
	\caption*{FIG. S8. The calculated phonon density of states at selected pressures.}
\end{figure}

\begin{figure}
	\centering
	\includegraphics[scale = 0.73]{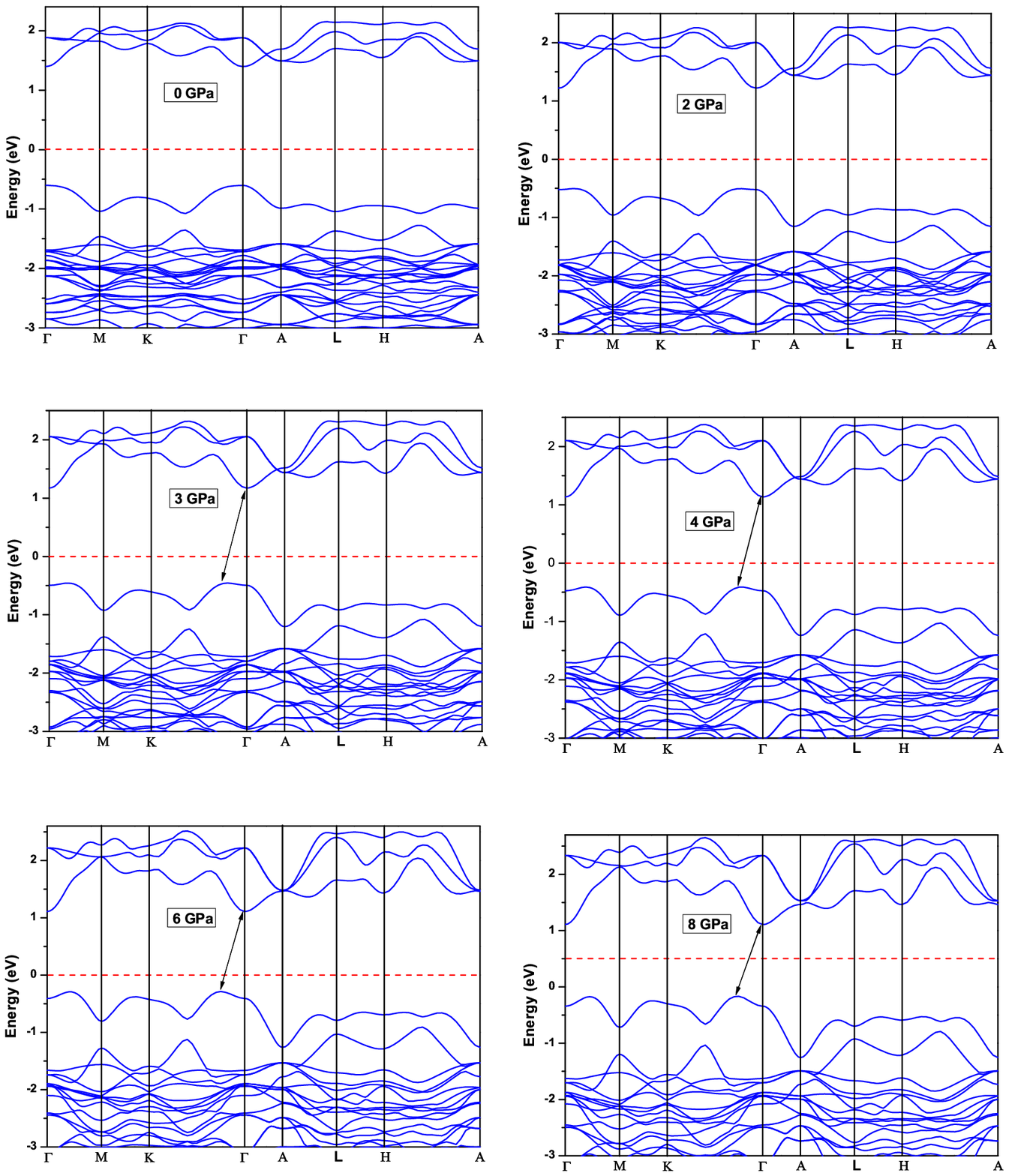}
	\vspace*{-50mm}
	\caption*{FIG. S9. The calculated electronic band structure at selected pressures. The diagram shows a direct-to-indirect band transition at 3 GPa.}
\end{figure}

\begin{figure}
	\centering
	\includegraphics[scale = 0.73]{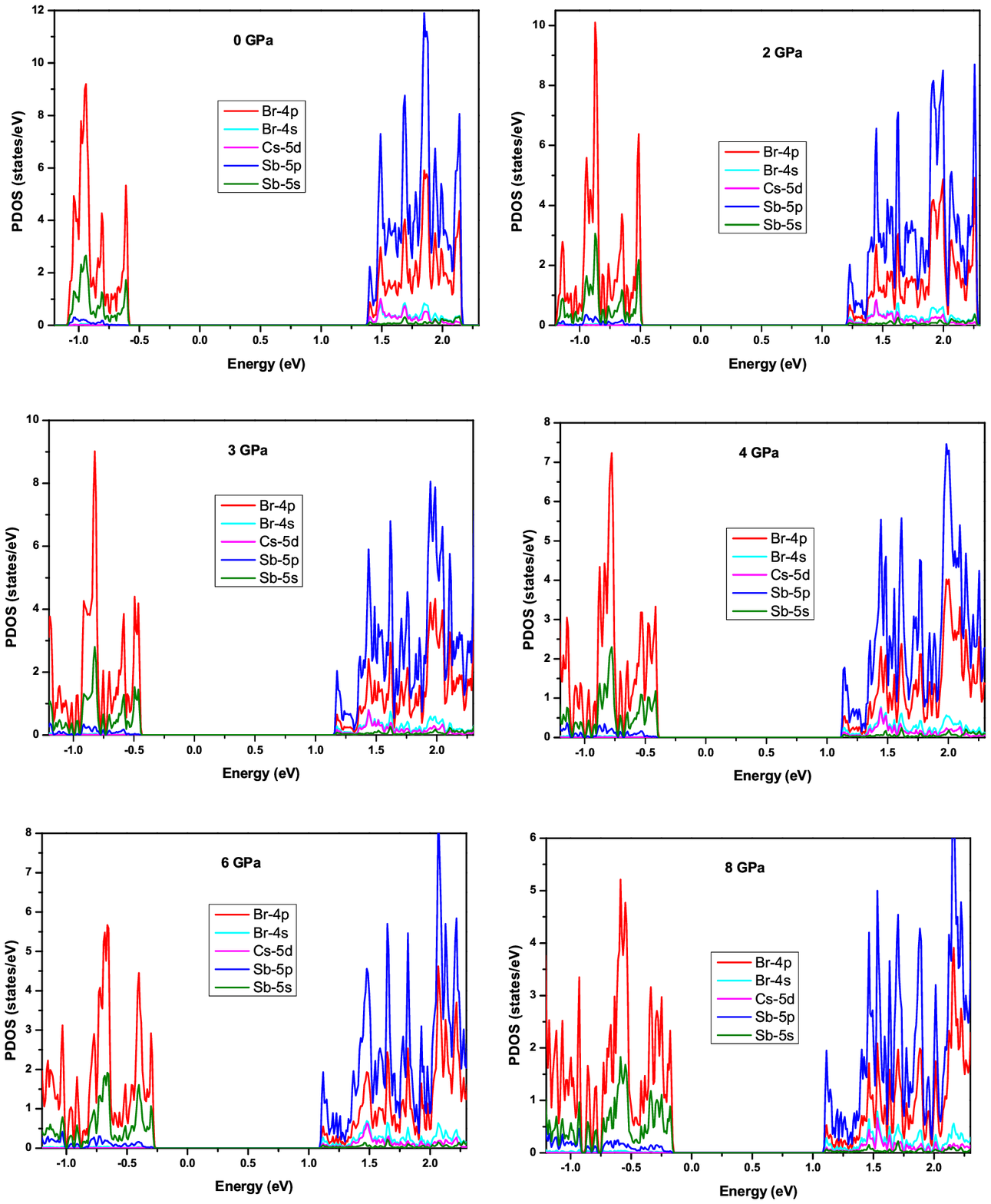}
	\vspace*{-50mm}
	\caption*{FIG. S10. The calculated projected density of states at various pressures, showing the contribution of different atomic orbitals in the conduction and valence band.}
\end{figure}